\documentclass{jfm}
\usepackage{graphicx}
\usepackage{epstopdf, epsfig}
\usepackage[colorlinks=true, citecolor=blue, linkcolor=blue]{hyperref}
\usepackage{amsmath}

\newcommand{\gm}{\gamma}

\newcommand{\df}{\mathrm{d}}

\newcommand{\Dlt}{\Delta}
\newcommand{\grav}{\mathrm{g}}

\shorttitle{Rotational flows beneath periodic waves near stagnation}
\shortauthor{L. Chen, B. Basu and C.-I. Martin}

\title{On rotational flows with discontinuous vorticity beneath steady water waves near stagnation}

\author{Lin Chen\aff{1,2},
    Biswajit Basu\aff{2} 
      \corresp{\email{basub@tcd.ie}}
      \and
      Calin-I. Martin\aff{3}
      }

\affiliation
{
\aff{1} Department of Bridge Engineering, Tongji University, Shanghai 200092, China
\aff{2} School of Engineering, Trinity College Dublin, Dublin 2, Ireland
\aff{3} Department of Mathematics, University of Vienna, Austria
}

\begin{document}

\maketitle

\begin{abstract}
We numerically investigate the flow structure of periodic steady water waves of fixed relative mass flux propagating on rotational flows with piece-wise constant vorticity. We show that for wave solutions along the global bifurcation diagram  the stagnation point can first occur internally or at the bottom, and then again occurs at the crest with further increase in wave height. We observe that the bifurcation diagram has a new branch which is not connected to the trivial solution. Furthermore, we present an in-depth discussion of the results on pressure distributions and particle trajectories beneath large-amplitude steady waves near stagnation. We also expand on previous results concerning the amplitude and mass flux of steady water waves traveling on rotational flows with discontinuous vorticity. 

\end{abstract}


\section{Introduction \label{sec:intro}}
The interaction between surface water waves and underlying currents affects the wave properties significantly \citep{longuet1961changes}, as also confirmed by experiments, e.g., \cite{swan2001experimental}. The interaction is hence of practical interest for design of offshore structures, see recent comparisons of structural responses with and without this interaction considered \citep{chen2018fatigue,chen2018wave} where linear wave-current interaction models were used though; for coastal engineering applications \citep{kirby1983propagation, dong2012theoretical} and 
in geophysical fluid dynamics \citep{kirby1984note, kirby1989surface}. Of most interest is the problem of two-dimensional (2D) periodic water waves propagating on steady current. A large number of theoretical studies and numerical investigations have been dedicated to this problem. However, it is still far from being fully understood. A comprehensive introduction to nonlinear wave-current interactions can be found in \cite{constantin2011nonlinear}. 

In fact, it is only recently that the existence of the large-amplitude wave solutions for rotational flows with general vorticity (thus allowing arbitrary current profiles) in finite water depth has been proved by \cite{constantin2004exact}. Prior to this, only waves of small-amplitude were known to exist in the presence of vorticity (\cite{dubreil1934sur}). \cite{hur2006global} further extended the proof of \cite{constantin2004exact} to the case of infinite-depth with the additional restriction that the vorticity distribution is positive and monotone. \cite{hur2011stokeswaves} later completed the proof of existence of large-amplitude waves with infinite depth removing the mentioned restrictions on vorticity. Moreover, \cite{constantin2011periodic} and \cite{strauss2012vorticity} addressed the existence of periodic traveling waves of small and large amplitude in a flow with an arbitrary bounded but discontinuous vorticity.

Following the work by \cite{constantin2004exact}, rotational flows with constant and non-constant vorticity have received considerable attention. Notably, many numerical investigations, on both irrotational and rotational flows, appeared much earlier. For example, for irrotational flow in finite water depth, \cite{chappelear1961direct} expanded the velocity field and the wave profile in Fourier series and the Fourier coefficients were numerically determined from the Bernoulli equation by using the least squares method. The direct numerical simulation results were compared with the fifth-order Stokes theory. \cite{dalrymple1974finite} presented a numerical model for steady water waves propagating over a linear shear current (constant vorticity) where stream function was expressed as the superposition of a linearly varying current profile and an irrotational wave field. The results thereof showed that the vorticity has a significant impact on the wave height. \cite{simmen1985steady} and \cite{da1988steep} studied periodic steady water waves with constant vorticity for infinite and finite water depth, respectively. The numerical methods for irrotational flows and rotational flows with constant vorticity are generally based on the boundary integral method or the Fourier expansion. Recently, \cite{clamond2018accurate} developed a novel conformal mapping method for irrotational flows, where the governing equation is rendered of Babenko kind which is then solved using the Petviashvili method. The model is very efficient and is able to solve for large amplitude waves with quite large steepnesses over arbitrary depth. \cite{dyachenko2019stokes} further modified the Babenko equation to permit constant vorticity and applied Newton-GMRES method following the use of the Fourier expansion. There are also other methods that have been developed. For instance, \cite{ablowitz2006new} proposed a non-local formulation which was further applied to consider non-flat bottom and moving non-flat bottom boundaries \citep{fokas2012water,francius2017two}. \cite{wahlen2009steady} and \cite{ehrnstrom2011steady} devised a novel transform for small-amplitude waves with constant and linear vorticity, i.e., a vertical scaling transformation. 

For waves on rotational flows with non-constant vorticity, early numerical investigations date back to \cite{dalrymple1973water,dalrymple1977numerical} where the author considered the effects of various current profiles on the finite-amplitude steady waves over finite water depth. The method was based on the Dubreil-Jacotin (DJ) transformation and finite difference discretization of the transformed rectangular domain. For a given boundary condition, the solution to internal points is improved by a number of sweeps; the error on the free surface boundary condition is then evaluated; subsequently, the solution on the free boundary is updated and the procedure is repeated until convergence. Using this method, \cite{dalrymple1976symmetric} showed the influence of the vorticity on wavelength and crest elevation of the wave, considering that the currents vary as trigonometric and hyperbolic sine and cosine functions of the depth. \cite{thomas1990wave} performed experimental studies for a finite-amplitude wave train interacting with a steady current containing an arbitrary distribution of vorticity, and moderate agreement was found between the measurements and the results obtained using the model of \cite{dalrymple1977numerical}.

\cite{ko2008large} presented a numerical continuation method to find large-amplitude 2D periodic steady water waves with arbitrary vorticity, which was used for discussing the relationship between amplitude, hydraulic head, depth and mass flux for variable vorticity cases \citep{ko2008effect}. This method neither uses truncated Fourier expansion nor makes any assumption on the water depth or wave amplitude. Further numerical analyses of water waves on rotational flows with discontinuous vorticity can be found in \cite{strauss2010steady}, \cite{constantin2011periodic}, and \cite{constantin2012dispersion}. The ethos of the approaches towards wave computation is that the local bifurcation from a laminar flow to a small-amplitude wave solution is predicted by a dispersion relation, the algorithm beginning with the linearized solution. The wave solution is then continued with increasing wave height until the solution becomes close to a wave in a flow with stagnation point (global bifurcation). The dispersion relation has been derived for small-amplitude periodic water waves traveling on rotational flows for various vorticity cases, e.g., see \cite{constantin2011periodic}; and \cite{constantin2012dispersion} and \cite{martin2015dispersion} for flows with two layers. \cite{karageorgis2012dispersion} discussed dispersion relations for several types of non-constant vorticities. High-order approximations for the constant vorticity case were discussed in \cite{constantin2015approximations}. \cite{amann2018numerical} expanded upon the method of \cite{ko2008large}, by allowing nonuniform grid points and enabling the continuation with vorticity as a bifurcation parameter. Using the latter scheme they successfully discovered a new part of the bifurcation curve for some critical constant vorticity values (see also \cite{kalimeris2018analytical}). Furthermore, \cite{constantin2015penalization} proposed a penalization method to solve for large-amplitude waves. Note that using the method of \cite{ko2008large}, the mean water depth is recovered from the solution, which leads to a small  variation in the value of mean water depth along the bifurcation curve \citep{ko2008effect}. The mean-depth can be held constant by following the reformulation proposed by  \cite{henry2013large,henry2013steady}. 

Considerable understanding of water waves on irrotational flows and flows with constant vorticity has been established recently. For irrotational flows, \cite{constantin2006trajectories} proved analytically that the particle trajectories are not closed on Stokes waves and \cite{henry2006trajectories} extended the proof to deep water. \cite{constantin2008particle} and \cite{matioc2010particle} showed that there are no closed particle orbits in linear periodic water waves. \cite{borluk2012particle} reconstructed the velocity fields and the particle paths of long-crested waves on irrotational flows using the KdV approximation. These particle paths were found to be consistent with the Stokes drift but with increasing amplitude. \cite{carter2020particle} studied the particle paths in the nonlinear Schr{\"o}dinger models. \cite{grue2014velocity} experimentally studied the velocity field below large-amplitude periodic water waves and the results are generally consistent with nonlinear theories. \cite{grue2017experimental} further investigated the particle paths in steep waves at finite water depth (without background current) experimentally, focusing on the Stokes drift. It was shown that for steep waves on irrotational flow, two closed particle paths occur at two vertical positions, in contrast to the theoretical results \citep{constantin2006trajectories}.  The lower closed particle paths occur at a position quite close to the bed due to the boundary layer effect, whereas the theoretical study by \cite{constantin2006trajectories} assumes inviscid water. It is also clear from \cite{grue2017experimental} that the thickness of the boundary layer is about 5 percent of the water depth, which is significantly affected by viscosity effect. \cite{paprota2018particle} developed a theoretical model for studying the particle paths of nonlinear waves generated in a closed flume and the Stokes drift velocity obtained from experimental data \citep{paprota2016experimental} were found to be quite consistent with theories. For a summary of the present analytical knowledge on the properties of irrotational flows, including the velocity field, pressure and particle trajectories, see \cite{constantin2015flow}. For waves on a linear-shear flow, \cite{ehrnstrom2008streamlines} studied the pattern of streamlines and particle trajectories and pointed out that for non-positive vorticity the particles display a forward drift. Using an asymptotic expansion for the stream function, \cite{ali2013reconstruction} constructed the pressure beneath steady long gravity water waves traveling on flows with constant background vorticity. \cite{curtis2018particle} investigated the effect of linear background shear on the properties of wavetrains, particularly on particle paths, in deep water using nonlinear Schrödinger models. Numerically, for understanding the particle paths or trajectories, \cite{nachbin2014boundary} proposed a method for particle trajectories in Stokes waves in the boundary integral formulation for irrotational and constant vorticity cases. A summary of the flow structure beneath rotational flows with constant vorticity and the corresponding numerical method can be found in \cite{nachbin2017capturing}. Lately, \cite{dyachenko2019stokesII} recovered various characteristics of wave profiles in flows with constant vorticity.


For rotational flow with general vorticity but subject to certain constraints, \cite{constantin2007rotational} studied the points of maximal horizontal velocity of waves near stagnation. Further understanding on the velocity field has been established by \cite{varvaruca2008some} and \cite{basu2019some}.  \cite{vasan2014pressure} discussed the pressure beneath rotational flows with constant vorticity pointing out interesting applications in wave profile recovery 
\citep{oliveras2012recovering,basu2017wave,basu2017estimation}. 

For surface waves on rotational flows, there are still a number of open problems yet to be solved \citep{strauss2012vorticity}. However, it becomes extremely difficult to capture the properties of the surface wave and the flow structure analytically for complex cases. As manifested in the literature, numerical investigations can provide us with important insights for
such cases. To deal with the otherwise intractable problem of rotational flows with arbitrary voriticity distribution, contrary to the widely assumed linear shear flows, we focus on the shear flow with piece-wise constant vorticity which is the simplest case deviating from constant vorticity. This is a valid approximation for waves traveling on depth-varying currents \citep{dalrymple1974water,swan2001experimental}. In particular, we are concerned with shear flows with two layers. It is known that wind typically induces vorticity in the layer near the surface, while near the ocean shore, vorticity can be generated at the bottom, e.g., by tidal action \citep{ko2008effect}. In this study, we report on a new branch of the bifurcation curve for waves on rotational flows with discontinuous vorticity and investigate the pressure beneath the waves for flows with nearly approaching stagnation points. Further, we develop a method for computing particle paths based on the exact velocity field solved numerically. The particle paths are of increasing interest since they are related to the Lagrangian transport and the Stokes drift in ocean  \citep{bremer2016lagrangian,van2017stokes}.

The rest of this paper is structured as follows. In Section \ref{sec:prob} and Section \ref{sec:method}, we briefly introduce the problem and the numerical methods respectively. The numerical results are then presented in Section \ref{sec:result}. We close the study by a brief summary in Section \ref{sec:con}.

\section{Statement of the problem}\label{sec:prob}
We consider two-dimensional steady, periodic, rotational gravity traveling surface waves, of period $2L$ ($L>0$ being a constant) propagating over water flows bounded below by the flat bed $y=-d$, where $d>0$ is a constant. The water is assumed to be inviscid. In the numerical investigation, we shall place our focus on the case with discontinuous vorticity of piece-wise constant type.

\subsection{Governing equations}
The time-dependent water wave problem is presented in the Cartesian $(X,Y)$-coordinate system with the origin on the mean water level, having the $X$-axis pointing in the horizontal direction and the $Y$-axis pointing upwards. The steady character of the problem we consider allows us to repeal the time-dependence by using the change of coordinates $(X-ct,Y)\mapsto (x,y)$, where $c$ is the constant speed of the traveling waves and $t$ denotes time. Hereafter, we will be working in the moving frame $(x,y)$ with the origin placed under the crest, except for the computation of particle paths. In this setting, the motion of the water flow is governed by the Euler (or momentum) equations, the equation of mass conservation supplemented by the kinematic boundary conditions on the two boundaries of the flow and the dynamic boundary condition on the free surface. The latter condition at the free surface decouples the motion of the water from the motion of the air above, while the former kinematic conditions convey the fact that the bed and the free surface are impermeable boundaries, cf. \cite{constantin2011nonlinear}.
 
Aiming at a simplification of the analysis we introduce the stream function, denoted by $\psi$, and defined (up to a constant) by 
\begin{equation}\label{eq:psi}
\psi_y=u-c\quad \mbox{and}\quad \psi_x=-v \ .
\end{equation}
Here, ($u,v$) is the velocity field in the $(x,y)$-coordinates.
The mass flux relative to the uniform flow at speed $c$ (normalized with respect to density) is 
\begin{equation}\label{p0}
p_0=\int_{-d}^{\eta(x)}[u(x,y)-c]\df y,
\end{equation}
further referred to as relative mass flux. It turns out from mass conservation equation and from the kinematic boundary condition on the surface that $p_0$ is a constant.

From \eqref{eq:psi} and the kinematic condition on the free surface we infer that $\psi$ is constant on the free surface.
Therefore, we choose $\psi=0$ on the surface and thus, by \eqref{p0}, we have
 $\psi=-p_0$ at the bottom $y=-d$. From the first relation in \eqref{eq:psi}, we have
\begin{equation}
\psi(x,y)
=-p_0+\int_{-d}^y\left[u(x,s)-c\right]\df s.
\end{equation}
Since we assume that $u<c$ throughout the fluid \citep{constantin2004exact}, $p_0$ is always negative.  Following \cite{constantin2004exact}, 
the fluid vorticity is defined as
\begin{equation}
\omega:=u_y-v_x.
\end{equation} 
Moreover, the assumption of no-stagnation (since $u<c$) implies, cf. \cite{constantin2011nonlinear}, that there is a function of one variable, which we denote by $\gamma$, such that $\omega(x,y)=\gamma(\psi(x,y))$ for all $(x,y)$ in the water domain. Also from formula \eqref{eq:psi}, we have that $\omega=\Delta\psi$. Hence, the fluid vorticity satisfies
\begin{equation}\label{eq:vort-def}
\Dlt\psi=u_y-v_x=\gm(\psi).
\end{equation}
Here, $\gamma$ is the function already found by the previous argument involving the no-stagnation condition, as presented in the book by \cite{constantin2011nonlinear}.

The density normalized hydraulic head $E$ is expressed as
\begin{equation}\label{eq:head}
E=\frac{1}{2}\left[(u-c)^2+v^2\right]+P+\grav y-\int_0^{\psi}\gm(s)\df s
\end{equation}
which is a constant owing to conservation of energy. In the expression, $P$ is the pressure (normalized with respect to density) in the fluid and at the surface $P=P_{atm}$, the atmospheric pressure which also normalized with respect to density. Therefore, for a given water depth, we can introduce another constant $Q=2(E-P_{atm}+\grav d)$, and the boundary condition on the surface is given by
\begin{equation}\label{eq:surf-eq}
\psi_x^2+\psi_y^2+2\grav(y+d)=Q\quad\mbox{on}\quad \psi=0.
\end{equation}

Eventually, we obtain the fixed domain boundary-value problem from the free-surface wave problem by applying the following DJ transformation, i.e.
\begin{equation}
q=x\quad\mbox{and}\quad p=-\psi(x,y)
\end{equation} 
to the relations \eqref{eq:vort-def} and \eqref{eq:surf-eq}, and including the boundary condition at $p=p_0$. The equations are
\begin{equation}\label{eq:eom}
\left.\begin{array}{lll}
(1+h_p^2)h_{pp}-2h_ph_qh_{qp}+h_{p}^2h_{qq}-\gm(-p)h_p^3=0 &\mbox{in}& R,\\
1+h_q^2+(2\grav h-Q)h_p^2=0 &\mbox{on}& p=0,\\
h=0 &\mbox{on}& p=p_0,
\end{array}\right\}
\end{equation}
where $h(q,p)=y+d$ is the height function and $R$ is the rectangle defined by $R=\left\{(q,p):-L<q<L,p_0<p<0\right\}$.

\subsection{Recovery of wave properties}
We will solve the problem \eqref{eq:eom} for the height function using a numerical method that will be introduced in the next section. For characterizing the waves, we shall recall the recovery of wave properties from the height function. Since in this formulation, the mean water depth $d$ is not fixed, it needs to be recovered first from $h$ by
\begin{equation}
d=\frac{1}{2L}\int_{-L}^{L}h(q,0)\df q \ .
\end{equation}
Afterwards, the water profile is determined by 
\begin{equation}
y=h-d.
\end{equation}
The wave height $a$ is obtained as
\begin{equation}
a=h(0,0)-h(L,0)
\end{equation}
as we are concerned with waves of one crest and one trough per period. Further, the velocity field is recovered by
\begin{equation}\label{eq:uv}
(c-u,v)=\left(\frac{1}{h_p},-\frac{h_q}{h_p}\right),
\end{equation}
recalling the DJ transformation (see \cite{constantin2011nonlinear} for details). The pressure in the fluid domain can be obtained from the relation \eqref{eq:head} and here we define a relative pressure, $\tilde P=P-P_{atm}$, for the following discussions as
\begin{equation}\label{eq:P}
\tilde P=-\frac{1+h_q^2}{2h^2_p}-\grav h+\frac{Q}{2}-\int_{0}^{p}\gm(-s)\df s = -\frac{\left[(u-c)^2+v^2\right]}{2}-\grav (y+d)+\frac{Q}{2}-\int_0^{p}\gm(-s)\df s.
\end{equation}

For computing water particle paths or trajectories, the traveling speed of the wave is recovered \citep{constantin2011nonlinear} by using
\begin{equation}\label{eq:speed}
c= \kappa-\frac{1}{L}\int_0^L \left[u(x,-d)-c\right]\df x,
\end{equation}
where $\kappa$ is the average horizontal current strength on the bed. We then need to solve a set of differential equations \citep{constantin2010pressure,nachbin2014boundary}
\begin{equation}\label{eq:pathode}
\left.\begin{array}{l}
X^\prime=u(X-ct,Y),\\
Y^\prime=v(X-ct,Y),\\
(X(0),Y(0))=(X_0,Y_0),
\end{array}
\right\}
\end{equation}
where the prime denotes the derivative with respect to time. Here, we use a numerical integration method to compute the particle paths once the velocity field is solved, as will be detailed in the next section.

\section{Numerical methods}\label{sec:method}
In this section, we briefly introduce the numerical methods to solve \eqref{eq:eom}, for performing numerical continuation to arrive at the solution of waves in flows with closely 
approaching stagnation points, and for computing the particle trajectories from the velocity field in the Eulerian setting. The dispersion relations and the linearized solutions for local bifurcation from the laminar flow solutions to small-amplitude wave solutions are also reviewed for the sake of completeness.

\subsection{Discretization and implementation}
Following the methods in \cite{ko2008large} and \cite{amann2018numerical}, we use a finite difference scheme to discretize the computational domain. Taking advantage of the symmetry of $h$ function with respect to $q=0$ \citep{constantin2007symmetry}, we set the computational domain as a fixed rectangle $[0,L]\times[p_0,0]$ to improve the computation efficiency. We first discretize the domain using uniformly distributed grid points and then refine the grid size near the surface and the bottom under the crest ($q=0$). In the case of discontinuous vorticity, we additionally refine the grid size near the location where the vorticity changes abruptly. We apply the central difference scheme at interior grid points, and at left and right boundary points by mirroring $h$ along $q=0$ and $q=L$, considering the periodicity and symmetry properties of the waves in absence of stagnation points. At the boundaries $p=0$ and $p=p_0$, we apply the backward and forward difference schemes, respectively. Note that the refinement is achieved by reducing the grid size smoothly so that the central difference still has a second-order accuracy, approximately.

For implementation, we use the open-source library \emph{Eigen} \citep{eigenweb} for solving the nonlinear algebraic equations subsequent to the finite difference discretization. We use the predictor-corrector method detailed in \cite{amann2018numerical} for numerical continuation and write our own code on the basis of \emph{Eigen}. In the following numerical analysis, we discretize the domain first by an equidistant $250\times750$ grid with additional refinement at locations where the stagnation point may occur.  Following the refinement, the smallest grid size achieved is ${1/8}^{\rm{th}}$ of the initial size prior to the mentioned refinement. 

\begin{figure}
	\centering
   \includegraphics[width=0.6\textwidth]{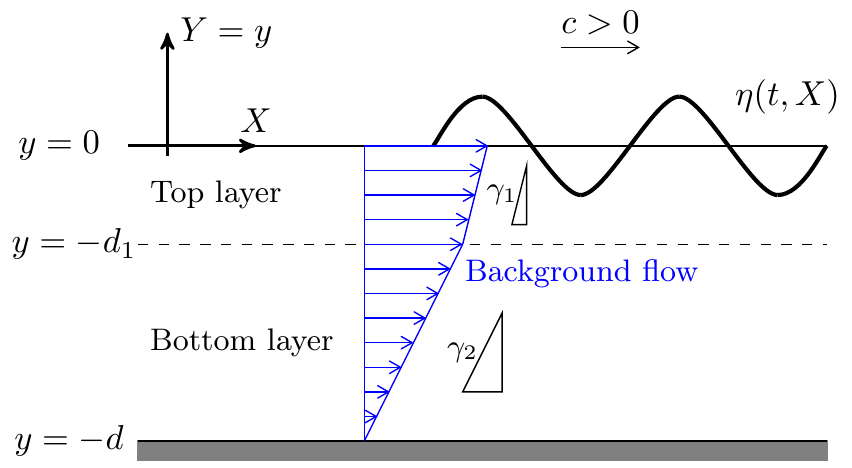}
   \caption{\raggedleft The setting of nonlinear periodic waves traveling on rotational flow with discontinuous vorticity. In the figure, $\gm_1$ and $\gm_2$ are positive. }
   \label{fig1:setting}
\end{figure}

\subsection{Dispersion relations and linearization solutions for local bifurcation}
As previously mentioned, we focus on the discontinuous vorticity. The vorticity function is expressed as
\begin{equation}
\gm=\left\{\begin{array}{ll}
\gm_1,&p\in(p_1,0),\\
\gm_2,&p\in(p_0,p_1).
\end{array}\right.
\end{equation}
where $p_1$ corresponds to the value of the streamline at the interface $y=y_1$, and $\gm_1$ and $\gm_2$ are the constant vorticity values of the surface layer and the layer adjacent to the bed. The setting of the problem is illustrated in Figure \ref{fig1:setting}. For the case of $\gm_1=\gm_2=0$, the problem reduces to irrotational flow, and the dispersion relation and the linearized solutions are available in \cite{constantin2011nonlinear}. For the situation of a rotational flow over an irrotational and vice versa, the dispersion relations are explicitly derived in  \cite{constantin2011periodic} and \cite{constantin2012dispersion}, respectively. For steady water waves on two rotational layers, see \cite{martin2015dispersion}. 

The dispersion relations are nonlinear equations. Here, we solve them using the Newton method. For large vorticity, we apply a simple continuation scheme by using the solution for smaller vorticity for initialization \citep{seydel2009practical}. 

\subsection{Computation of the particle trajectories}
\begin{figure}
	\centering
   \includegraphics[width=0.7\textwidth]{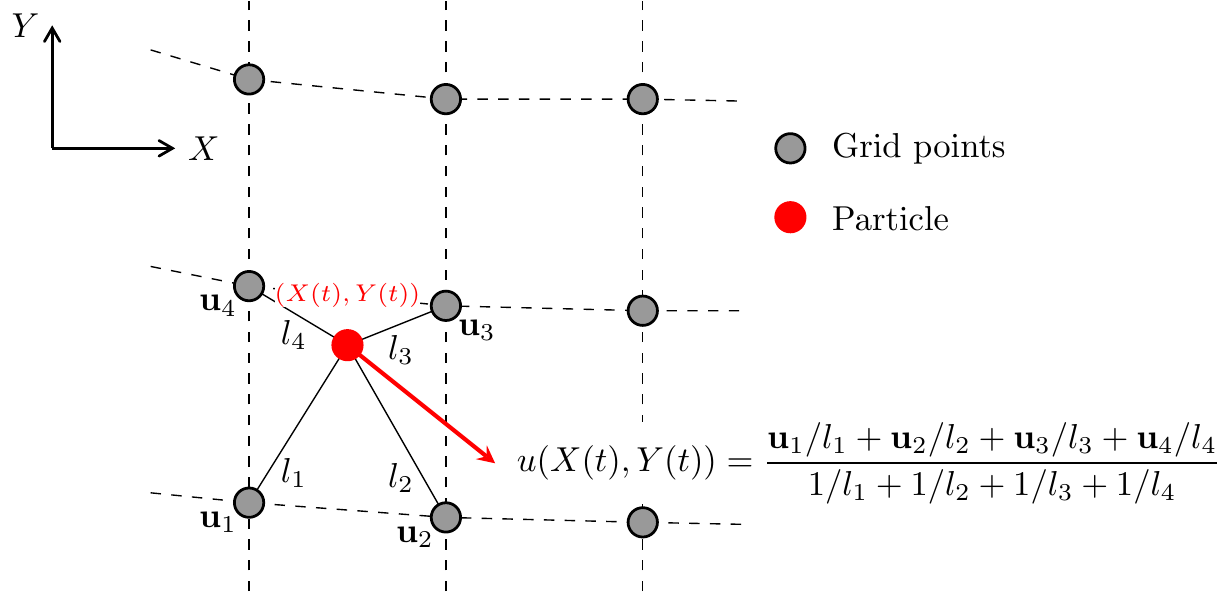}
   \caption{Interpolation strategy for instantaneous particle velocity. In the figure, $\mathbf{u}=[u\ \ v]^\top$ is the velocity vector; $l_1, l_2, l_3$ and $l_4$ are respective distances of the particle to the four neighboring grid points.}
   \label{fig2:interpolation}
\end{figure}

We numerically compute the particle paths in the physical domain, i.e. in the fixed $(X,Y)$-coordinate. For a particle located at $(X(t),Y(t))$ at time $t$, its velocity is interpolated from the velocities at the neighboring four grid points (following appropriate translation) using a similar interpolation method as in \cite{umeyama2012eulerian}. The interpolation scheme is illustrated in figure \ref{fig2:interpolation}. Note that in our case, the cell is of trapezoidal shape in the physical domain.  Subsequently, we use numerical integration to solve the first-order differential equations \eqref{eq:pathode}. This simple interpolation scheme is found to be sufficiently accurate.

\section{Results}\label{sec:result}
Following \cite{ko2008large,ko2008effect} and \cite{amann2018numerical}, we assign $\grav=9.8$, $2L=2\upi$ and $p_0=-2$. We focus on the new part of the bifurcation curves, pressure distribution, and particle paths under flows with nearly approaching stagnation points. Note that in the present system all the variables are expressed in SI unit and a physical system of a different wavelength can be scaled to the $2\pi$-periodic system analyzed in this paper by scaling $\grav$ and $\gm$ using the wave number \citep{constantin2011nonlinear}. However, the scaling has no influence on the laminar flow solution. In the following, as will be seen, the current velocity at the crest (under zero-amplitude waves) in most of the cases is less than 2.5 m/s, which is physically realistic in oceans. 

\subsection{Critical vorticity and the new part of the branch of bifurcation diagram}
For the case of constant vorticity, it is known that when the vorticity is below a critical value $\gm_{crit}$, the stagnation point first appears at the bottom and at that stage numerical continuation breaks down \citep{ko2008large}. However, \cite{amann2018numerical} and \cite{kalimeris2018analytical} found that when $\gm<\gm_{crit}$ but is very close to 
$\gm_{crit}$, the branch of the bifurcation diagram ($a-Q$ plot) has another part which is bounded below and above by two values of the bifurcation parameter $Q$. These bounds
correspond to two different waves in flows with stagnation point at the bottom and at the surface. On the new part of the branch, the wave height is further increased, and it is thus referred to as the upper part. To reach this upper part of the solutions, we need to overcome the gap of the bifurcation curve corresponding to waves in flows with stagnation points. For this purpose, \cite{amann2018numerical} resorted to the $\gm-$continuation scheme where a solution on the bifurcation diagram corresponding to $\gm_0$ ($\gm_0>\gm_{crit}>\gm$) is used to initialize local bifurcation to arrive at one solution on the upper part corresponding to $\gm$. Once one solution is obtained, the continuation in the solution set corresponding to $\gm$ can be performed again with $Q$ or the wave height as the bifurcation parameter. In the following, we apply this scheme when needed.

\begin{figure}
   \centering$\begin{array}{cc}
   \includegraphics[trim={0.4cm 0.8cm 0 0},clip,height=3.8cm]{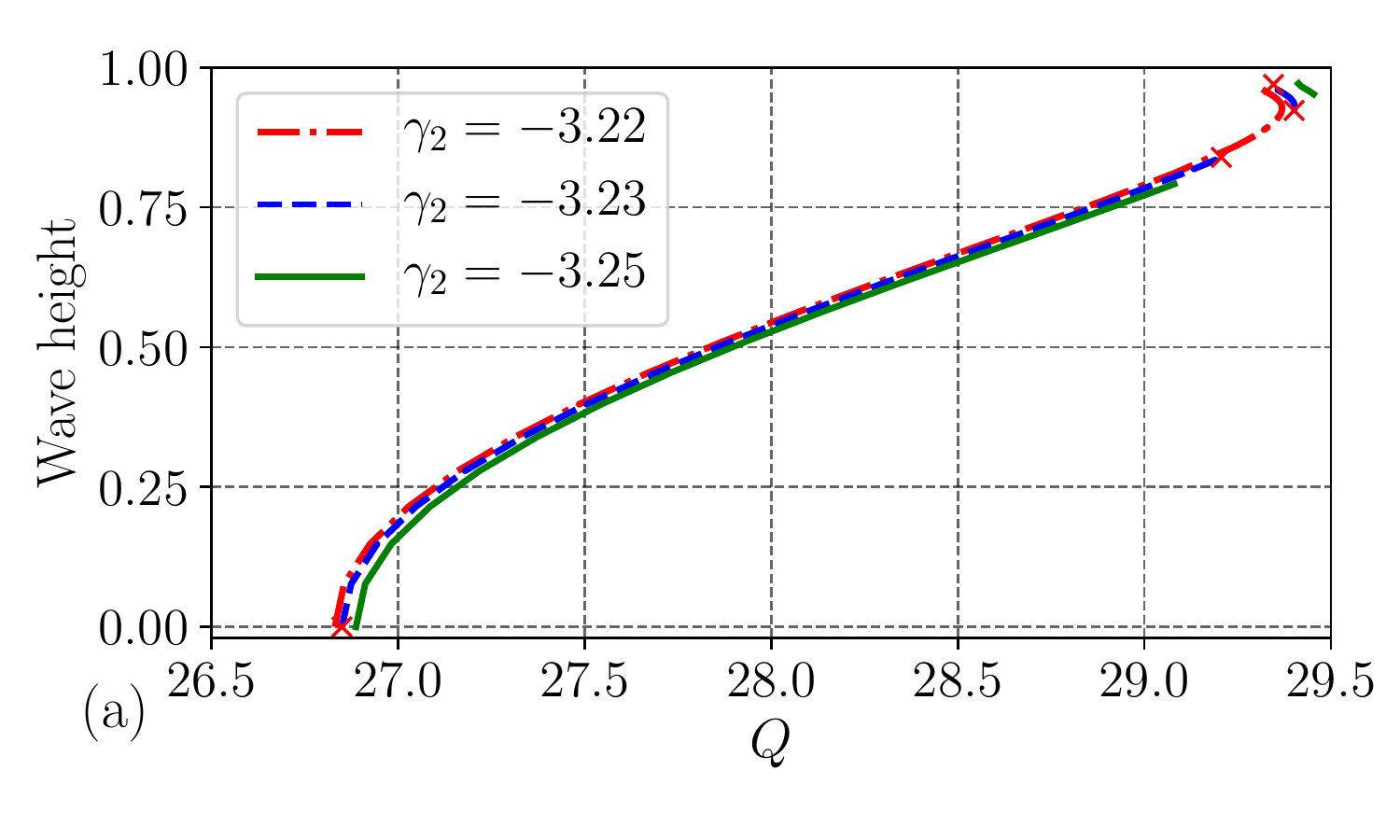} &
   \includegraphics[trim={0.6cm 0.8cm 0 0},clip,height=3.8cm]{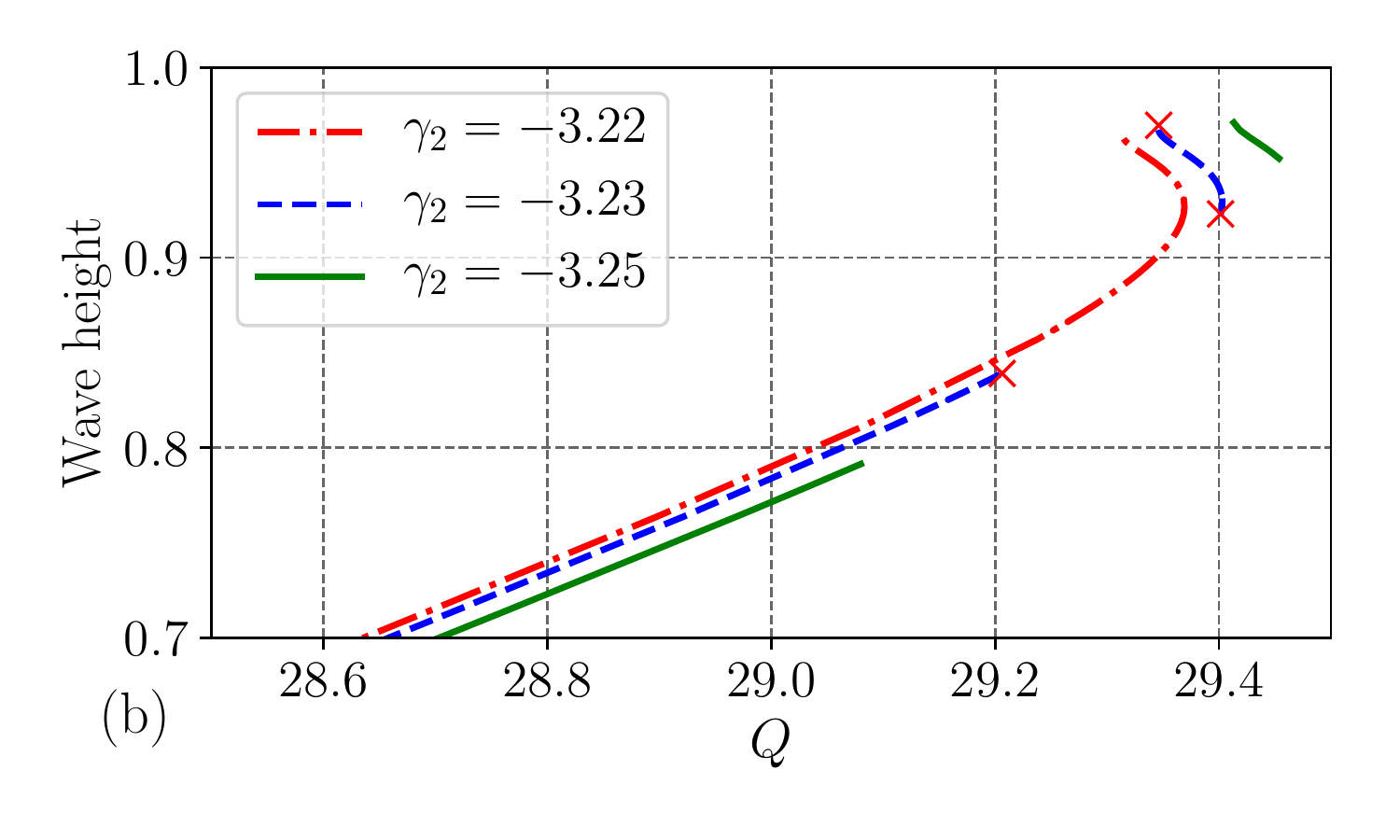} \\
   \end{array}$
   \caption{Bifurcation diagrams for cases with $p_1=-0.5, \gm_1=0$ (at the top layer) and various different values of $\gm_2$ (at the bottom layer): (a) full diagrams, and (b) a close-up view of the upper parts.}
   \label{fig3:bifurcation}
\end{figure}

\begin{figure}
   \centering$\begin{array}{cc}
   \includegraphics[trim={0.6cm 0.6cm 0 0},clip,height=3.8cm]{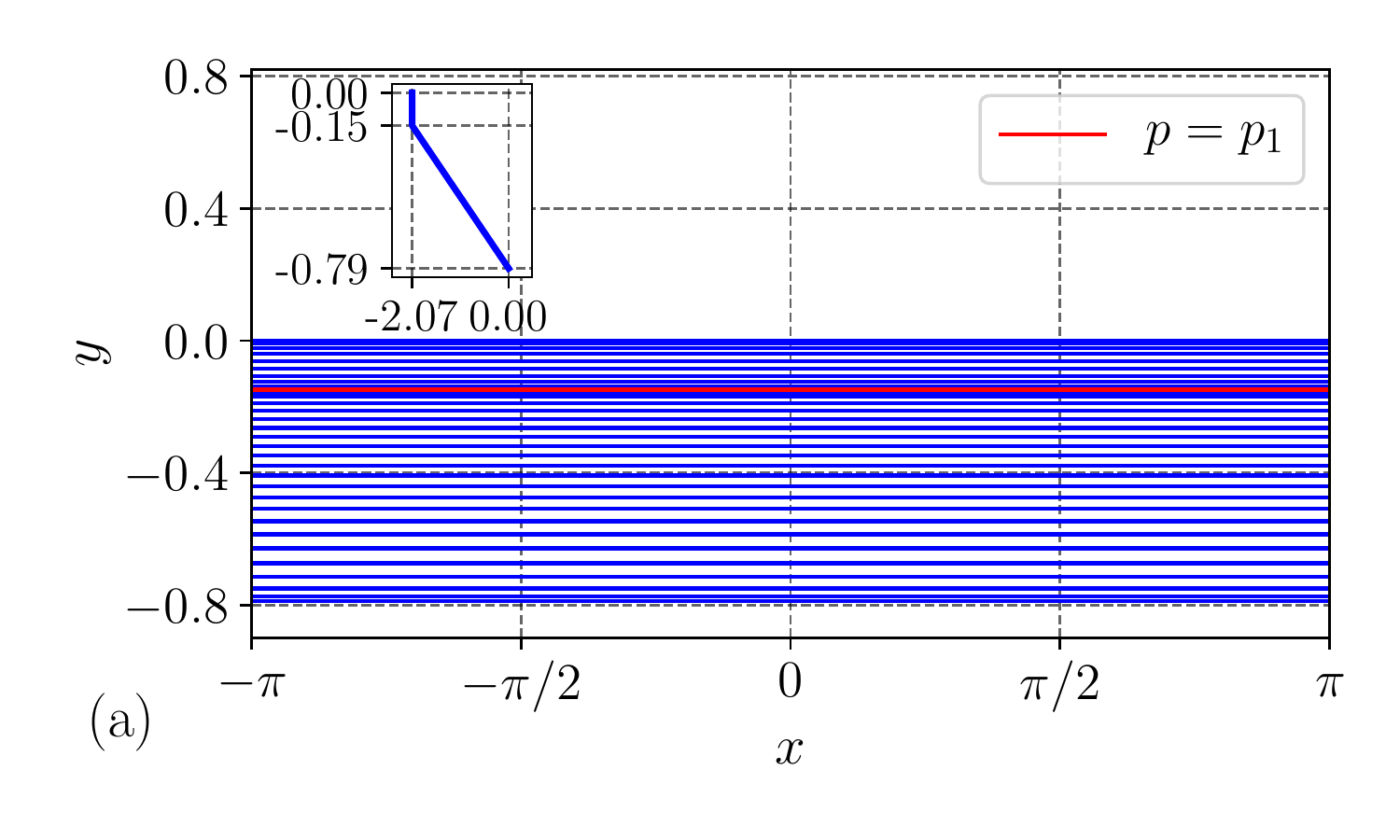} &
   \includegraphics[trim={0.6cm 0.6cm 0 0},clip,height=3.8cm]{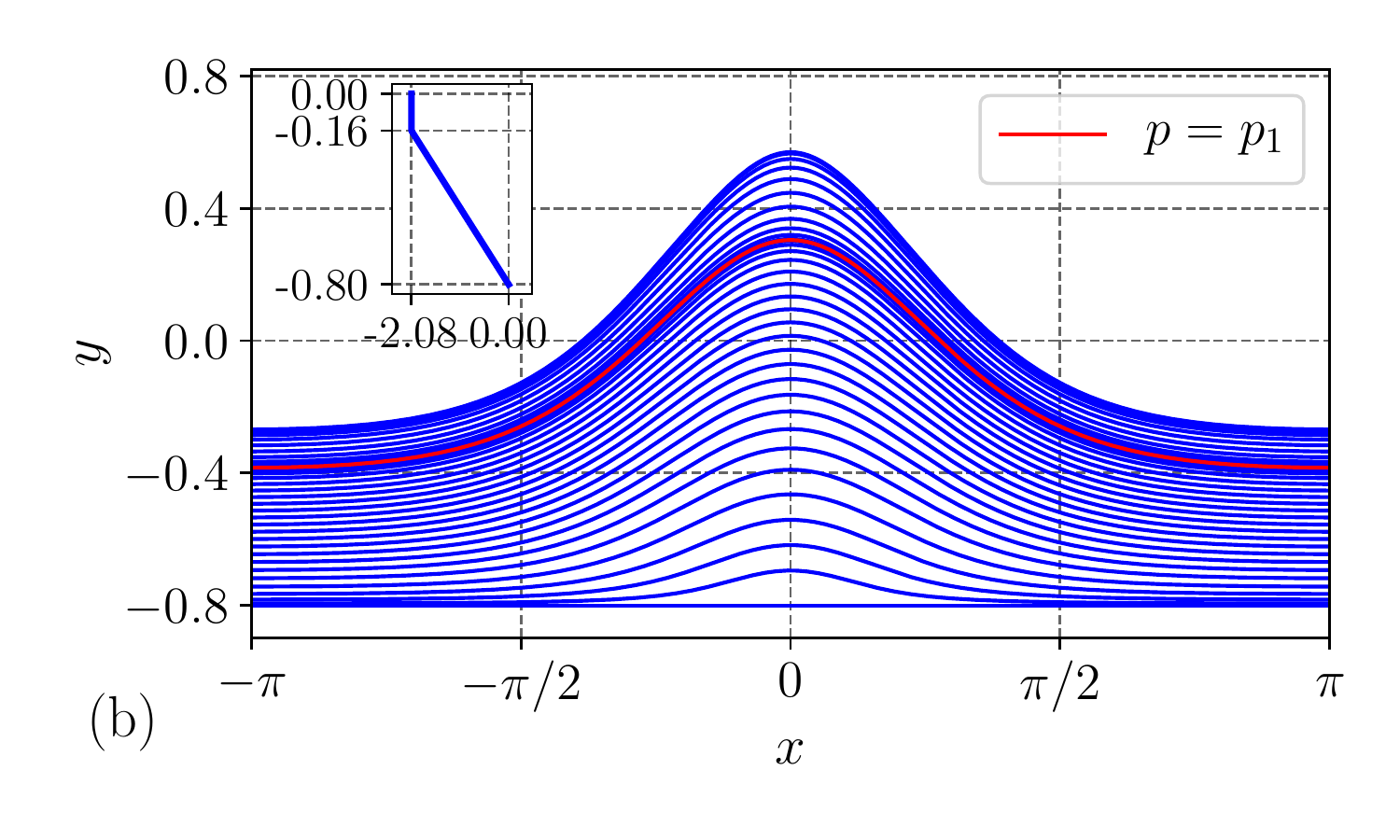} \\
   \includegraphics[trim={0.6cm 0.6cm 0 0},clip,height=3.8cm]{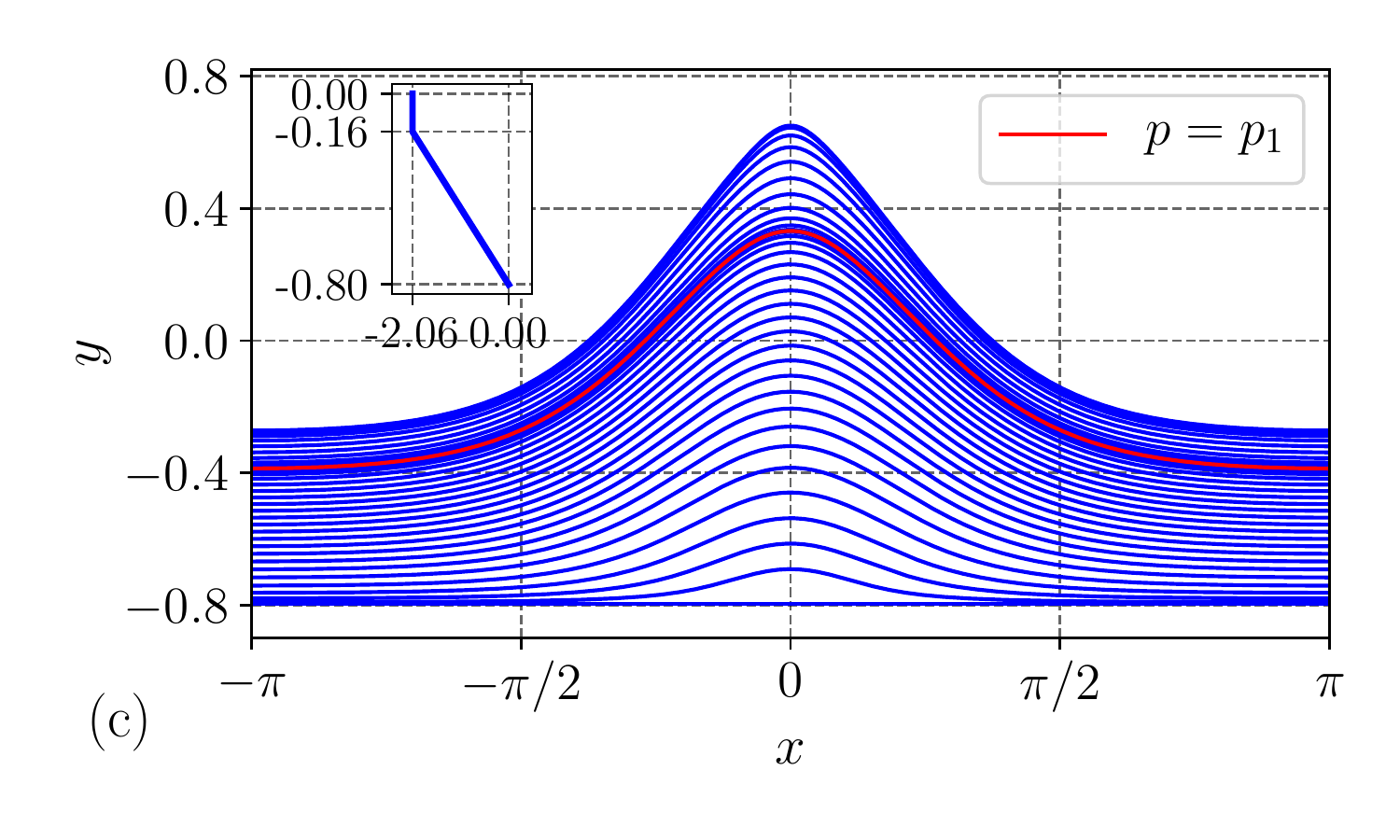} &
   \includegraphics[trim={0.6cm 0.6cm 0 0},clip,height=3.8cm]{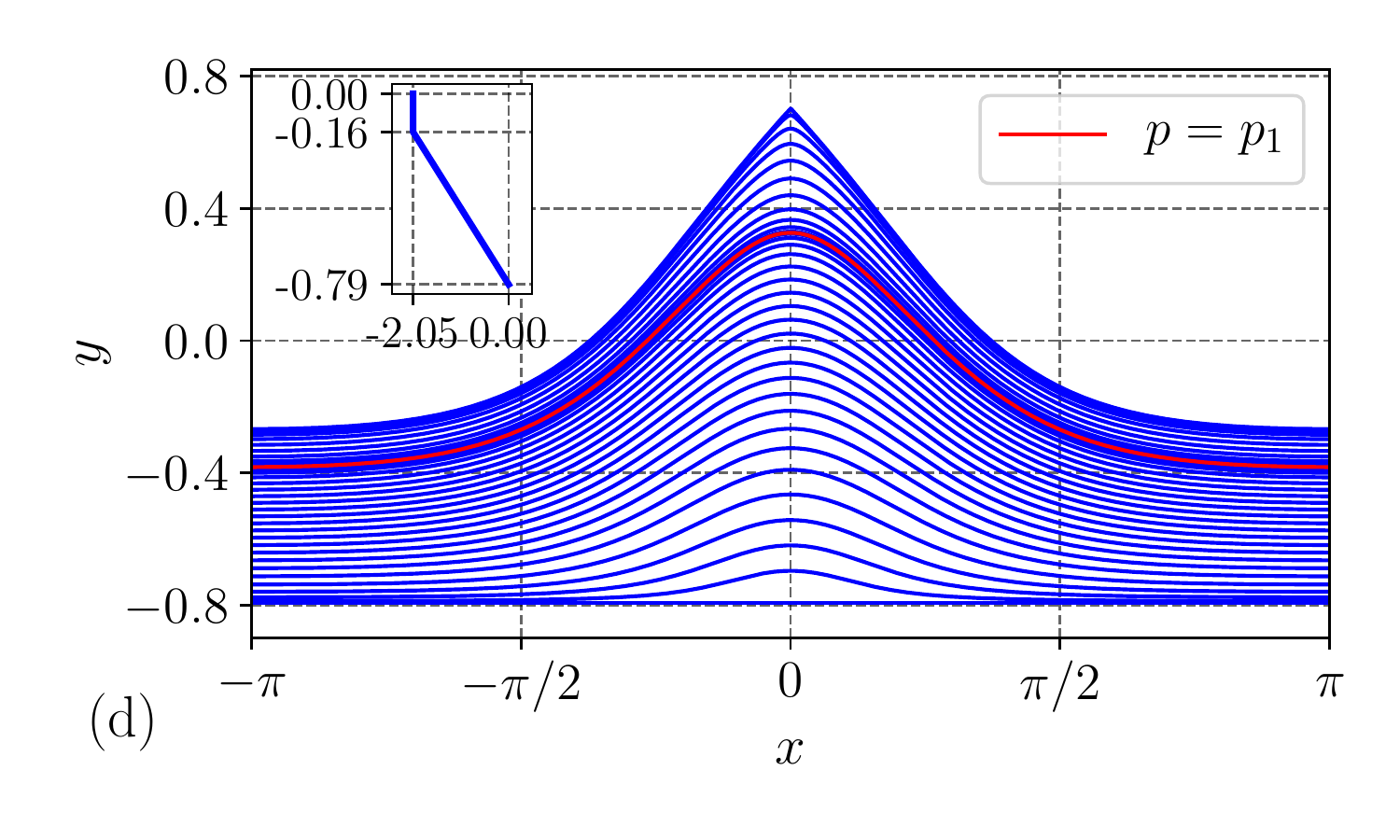} \\
   \end{array}$
   \caption{Streamlines of flows at the endpoints of each of the two parts of the bifurcation curve corresponding to $p_1=-0.5, \gm_1=0$ (at the top layer), and $\gm_2=-3.23$ (at the bottom layer): (a) laminar flow, (b) wave at the upper bound of the lower part, (c) wave at the lower bound of the upper part, and (d) wave at the upper bound of the upper part. In the figures, the insets show the background current profile when no wave motion.}
   \label{fig4:streamlines}
\end{figure}

\begin{figure}
   \centering$\begin{array}{cc}
   \includegraphics[trim={0.8cm 0.8cm 0 0},clip,height=3.8cm]{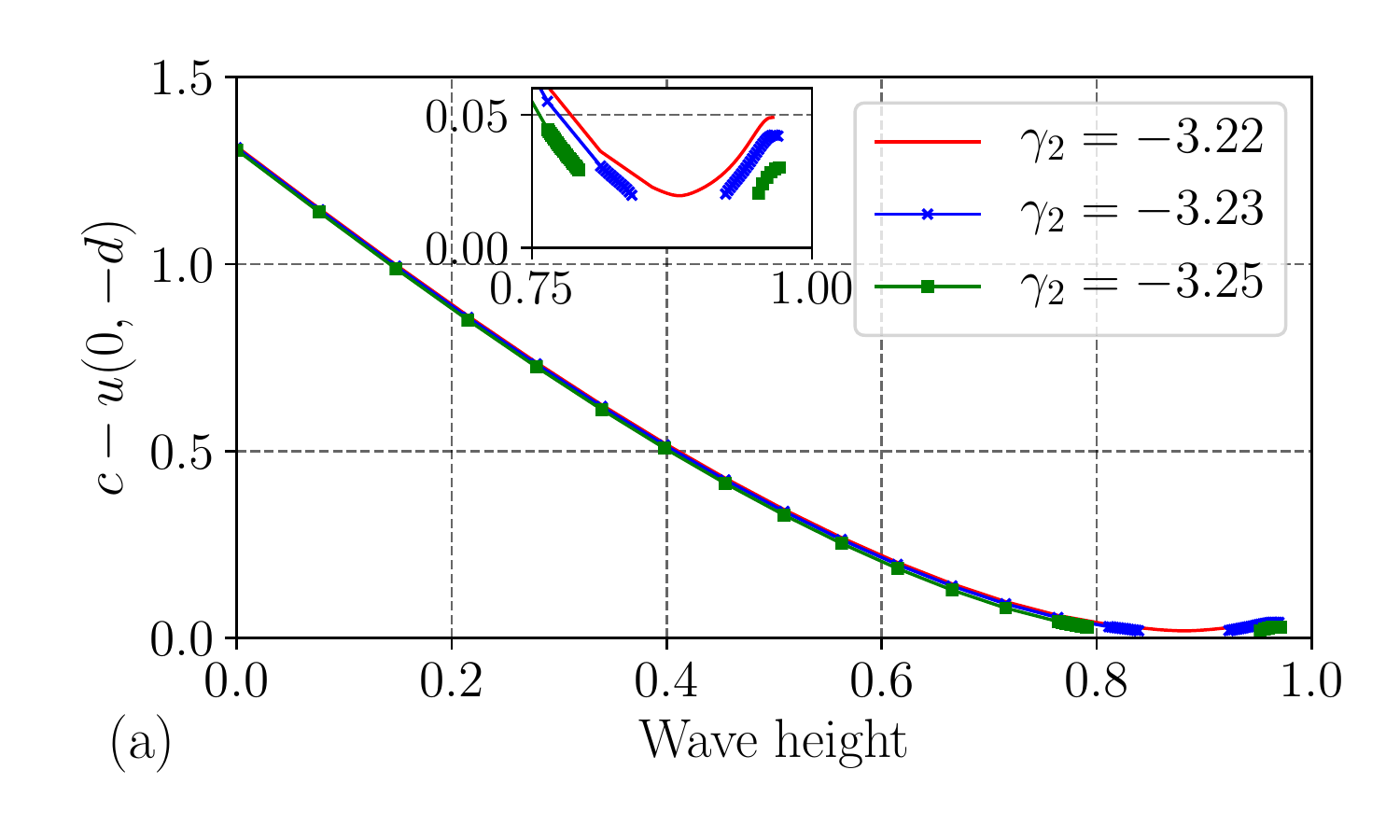} &
   \includegraphics[trim={0.7cm 0.8cm 0 0},clip,height=3.8cm]{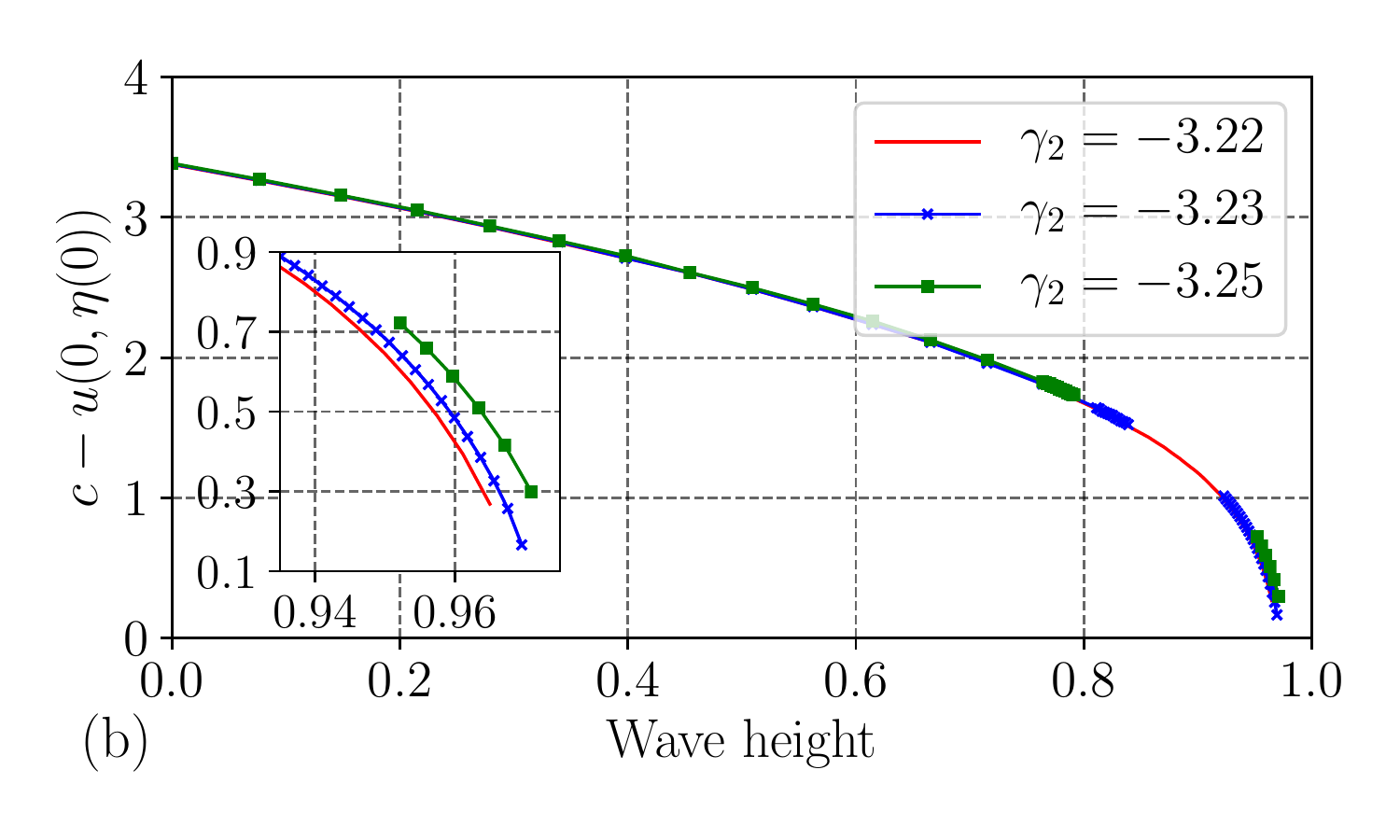} \\
   \end{array}$
   \caption{Horizontal velocity, $c-u$, under the crest along the bifurcation curve for $\gm_1=0$ (at the top layer), $p_1=-0.5$, and various different values of $\gm_2$ (at the bottom layer): (a) at the bottom and (b) at the surface. In the figures, the insets are closeup view of the curves when approaching the largest wave height.}
   \label{fig5:velocity}
\end{figure}

Proceeding from the observations of \cite{amann2018numerical}, we investigated the case of discontinuous vorticity and found that the upper part also exists. Furthermore, for the case of discontinuous vorticity, the upper bound on this part always corresponds to waves in flows with the stagnation point at the surface, while its lower bound corresponds to waves in flows with stagnation point either at the bottom or at an internal point where vorticity jump occurs. We first show an example of the former situation. An irrotational flow ($\gm_1=0, p_1=-0.5$) was considered on a rotational flow with vorticity $\gm_2$. We varied $\gm_2$ to find out the critical value $\gm_{2,crit}$ below which the stagnation point first occurs at the bottom. It was found that $\gm_{2,crit}$ is slightly less than $-3.22$. Note that with the vorticity values $\gm_1=0$ and $\gm_2=-3.23$ and hence a shearing laminar flow with profile shown in figure 4a over a total water depth of 0.787 m, we have the surface velocity, $u=-2.07$ m/s (the current flow direction is opposite to the wave propagation direction). Eq. (2.5) with a bifurcation value of $Q = 26.85$ 
(see figure 3a) leads to the relative surface velocity with respect to the wave speed as $c-u=3.38$ m/s, since $u<c$. The wave speed $c$ is $1.31$ m/s as also verified by the dispersion relation for this case. In the case of positive vorticity in the bottom layer, e.g., $\gm_1=0$ and $\gm_2=3$, the water depth is 0.74 m, the current velocity at the surface is around 1.53 m/s, and at the local bifurcation the wave speed is $c=3.69$ m/s. These values of the physical parameters indicate that the setting considered in this paper are realistic.

We then further decreased $\gm_2$ and found the upper part, as shown in figure \ref{fig3:bifurcation}. For the case of $\gm_2=-3.23$, the streamlines of wave solutions at the four endpoints (marked with cross in figure \ref{fig3:bifurcation}) of the two parts of the bifurcation curve are shown in figure \ref{fig4:streamlines}. Figure \ref{fig4:streamlines} clearly shows that the stagnation point first occurs at the bottom and then at the crest. Figure \ref{fig5:velocity} shows the horizontal velocities $c-u$ at the bottom and at the surface under the crest along the bifurcation curves. Note that both waves in figure \ref{fig4:streamlines}b and figure \ref{fig4:streamlines}c correspond to flows close to having a stagnation point at the bottom, and the horizontal velocities $c-u$ at the bottom in these two cases are comparable, as in figure \ref{fig5:velocity}a, while the wave in figure  \ref{fig4:streamlines}c has a larger height, a sharper crest and smaller horizontal velocity $c-u$ at the surface (see figure \ref{fig5:velocity}b). It is seen from figure \ref{fig5:velocity} that the velocity $c-u$ at the bottom is first decreasing and then increasing along the bifurcation curve, with the presence of a gap for $\gm_2<\gm_{2,crit}$, while the velocity $c-u$ at the crest is decreasing monotonically. Indeed, at the uppermost end of the bifurcation diagram in all those three cases, the velocities $c-u$ at the surface are still larger than those at the bottom, the stagnation point is expected to occur at the crest because the velocity at the surface is decreasing while at the bottom is increasing with respect to the increasing wave height.

\begin{figure}
   \centering$\begin{array}{cc}
   \includegraphics[trim={0.75cm 0.8cm 0 0},clip,height=3.8cm]{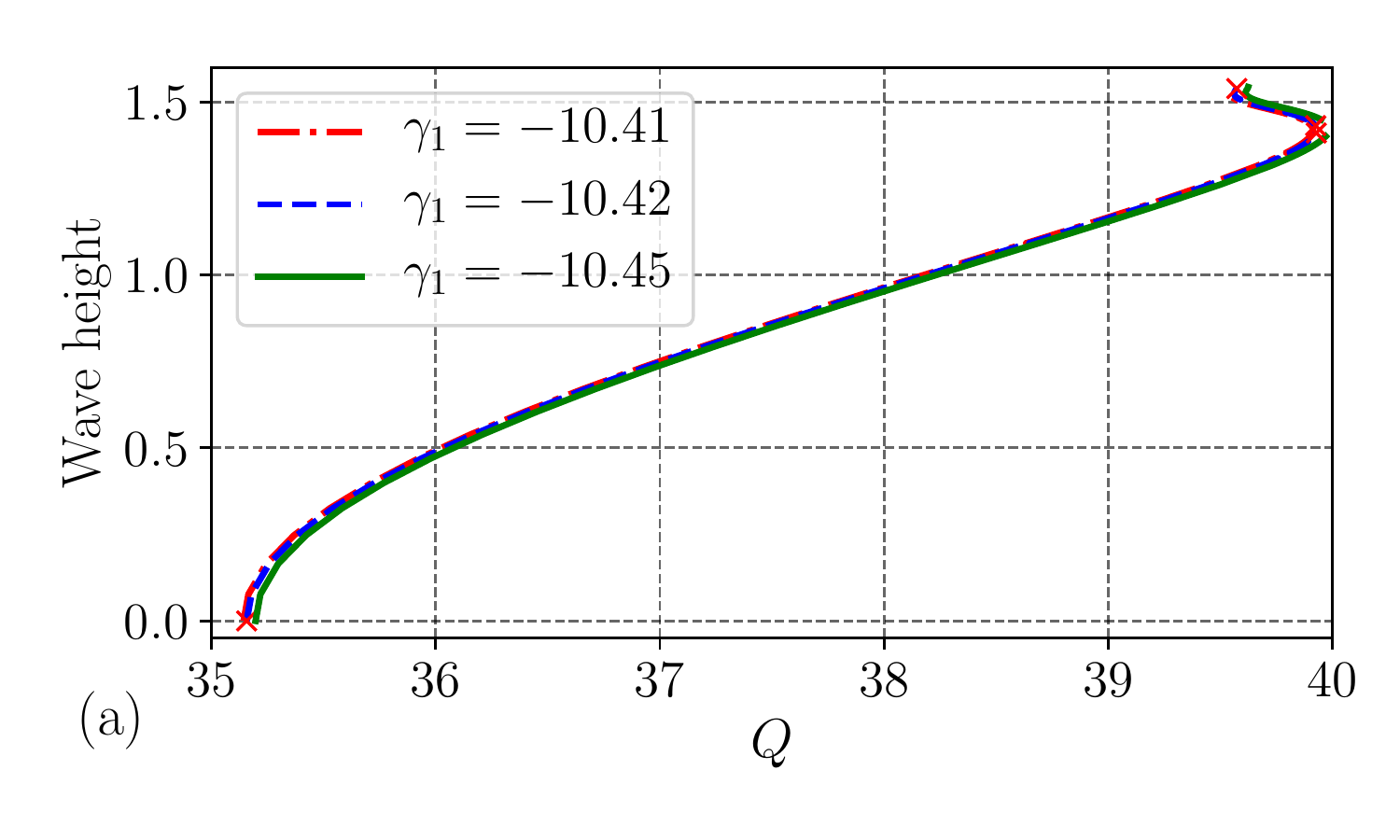} &
   \includegraphics[trim={0.75cm 0.8cm 0 0},clip,height=3.8cm]{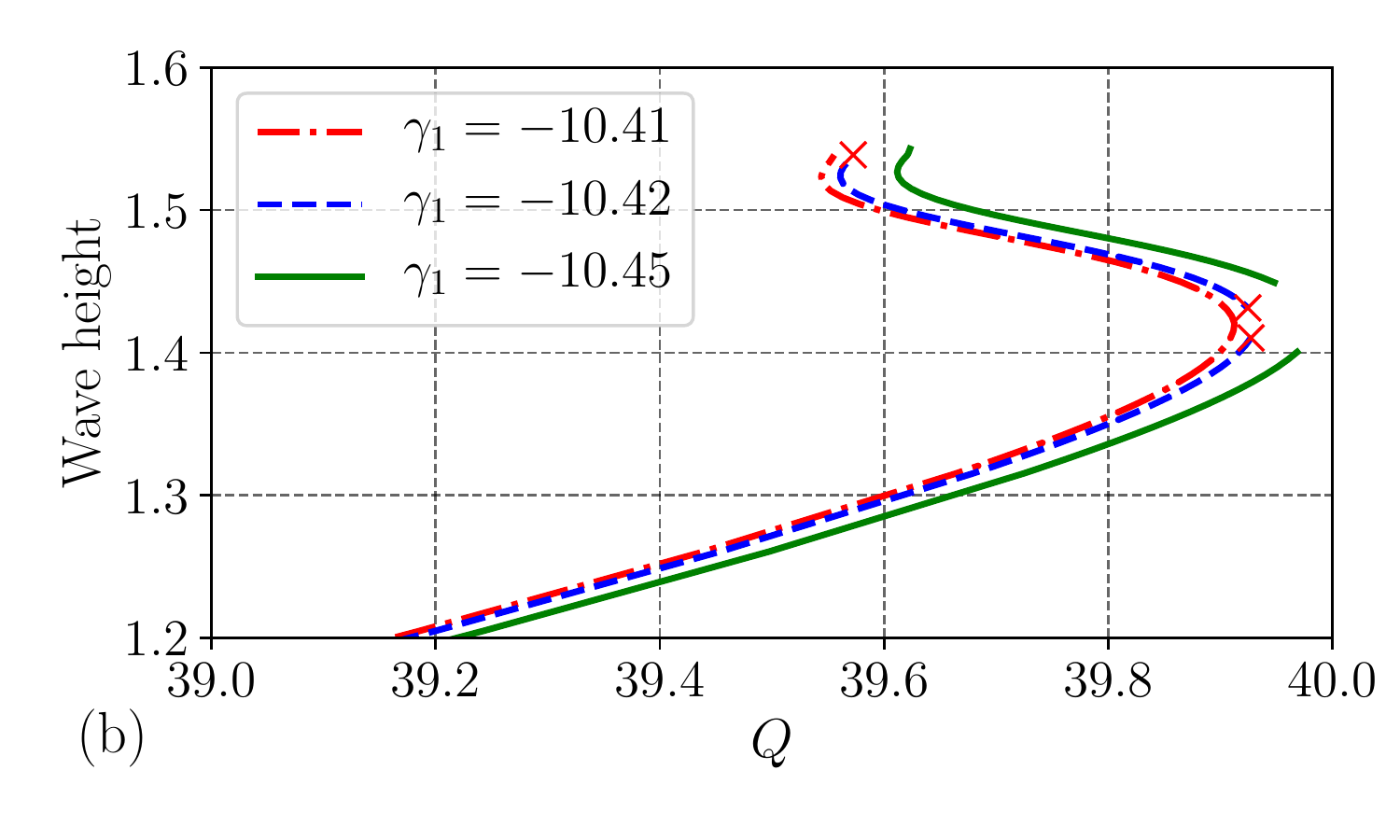} \\
   \end{array}$
   \caption{Bifurcation diagrams for cases with $p_1=-0.7, \gm_2=0$ (at the bottom layer) and various different values of $\gm_1$ (at the top layer): (a) full diagrams, and (b) a close-up view of the upper parts.}
   \label{fig6:bifurcation}
\end{figure}

\begin{figure}
   \centering$\begin{array}{cc}
   \includegraphics[trim={0.6cm 0.6cm 0 0},clip,height=3.8cm]{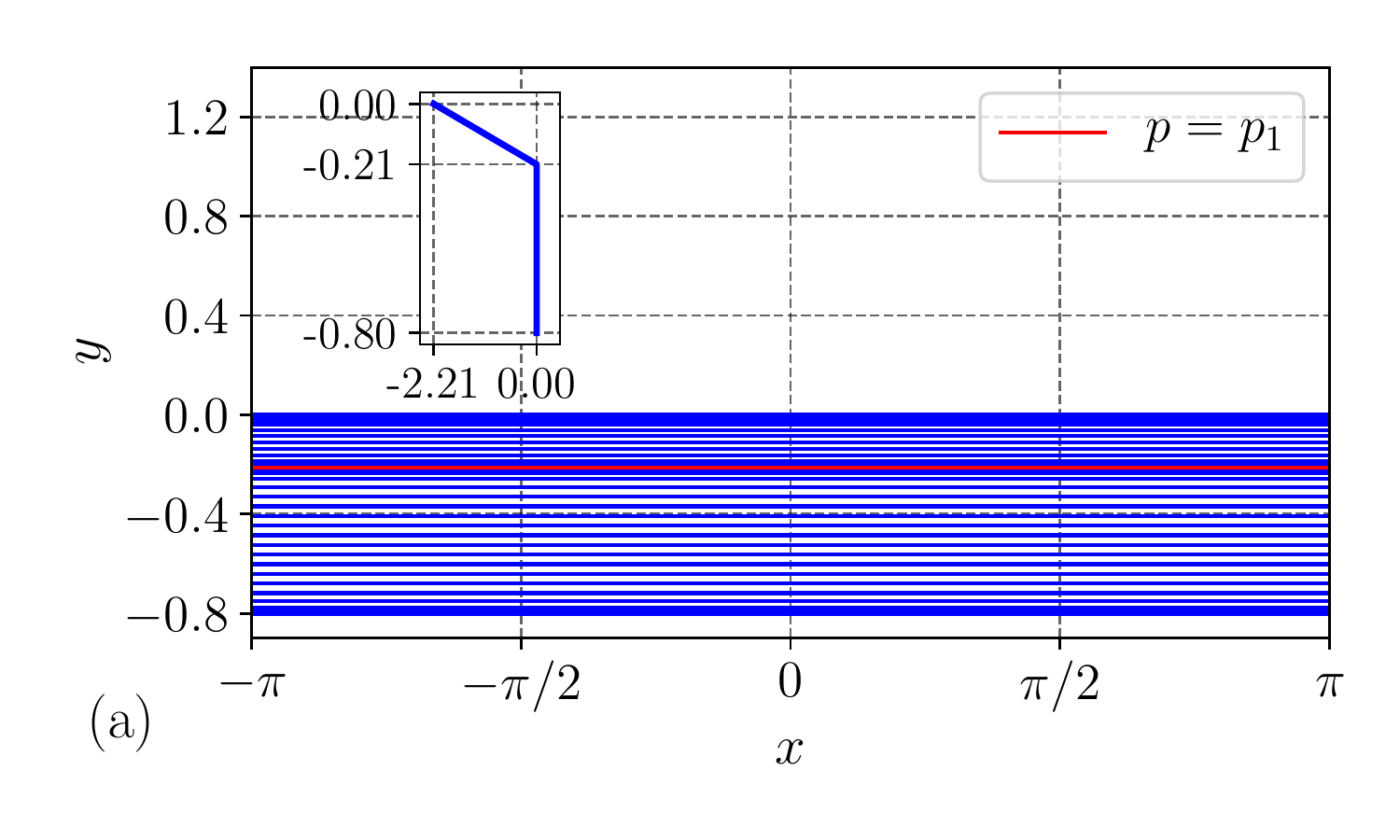} &
   \includegraphics[trim={0.6cm 0.6cm 0 0},clip,height=3.8cm]{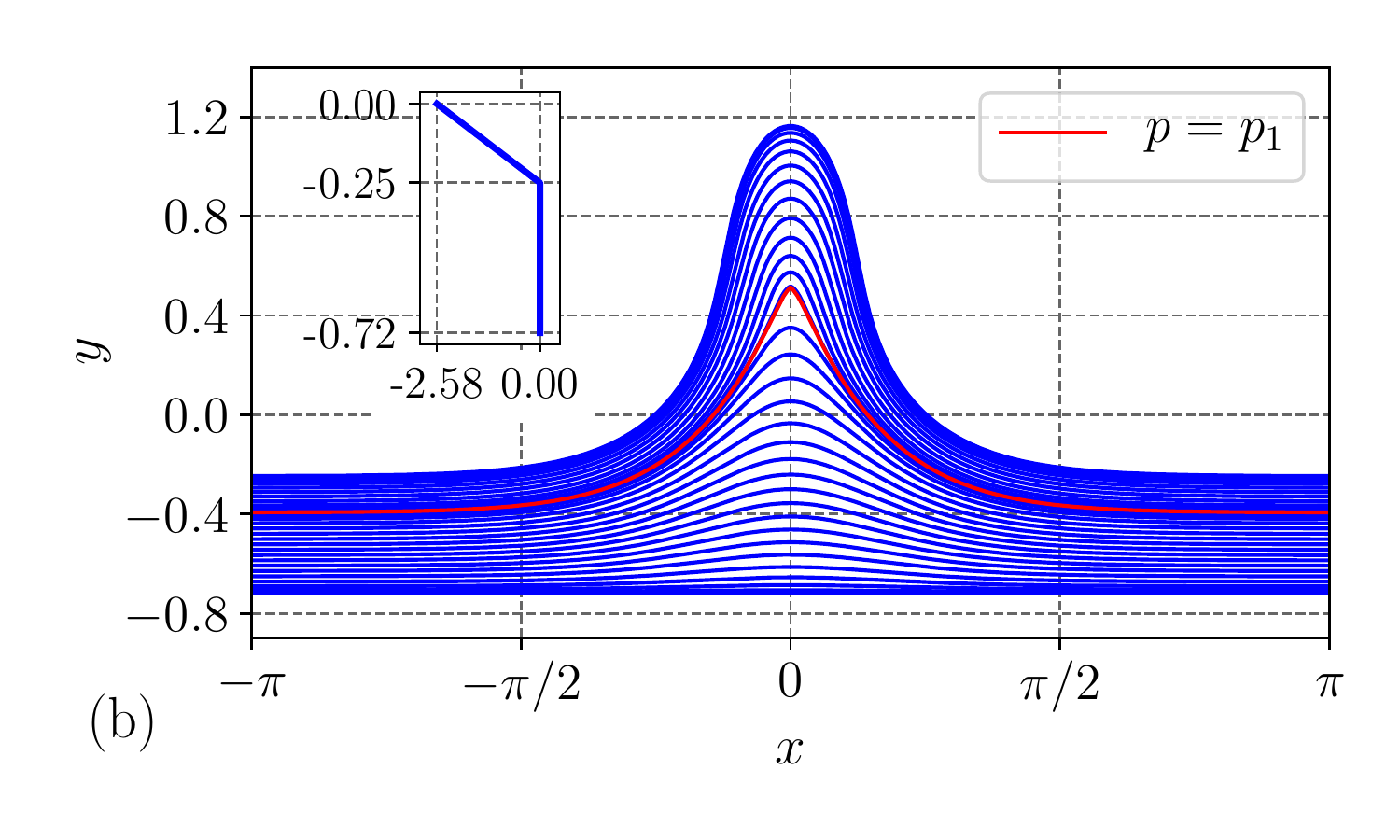} \\
   \includegraphics[trim={0.6cm 0.6cm 0 0},clip,height=3.8cm]{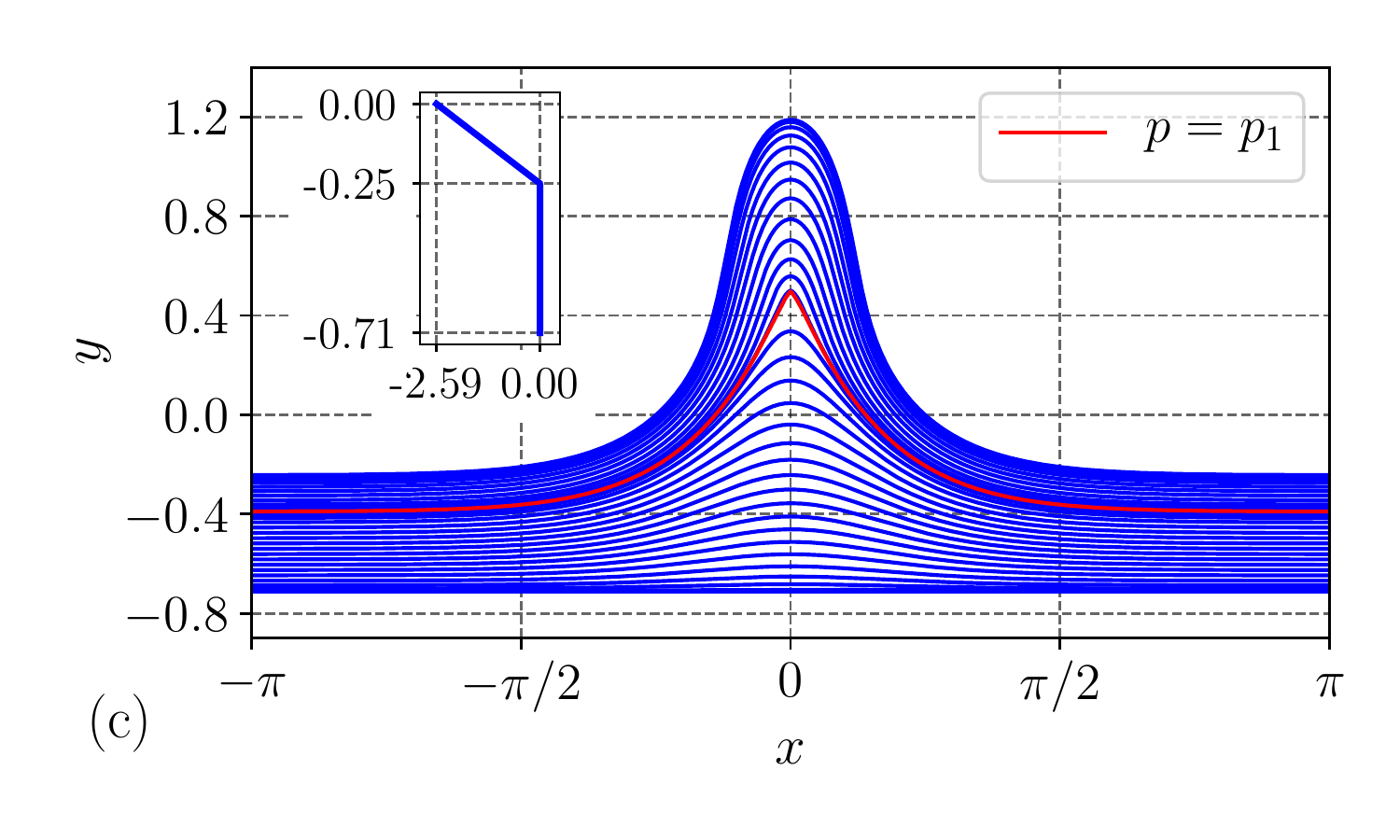} &
   \includegraphics[trim={0.6cm 0.6cm 0 0},clip,height=3.8cm]{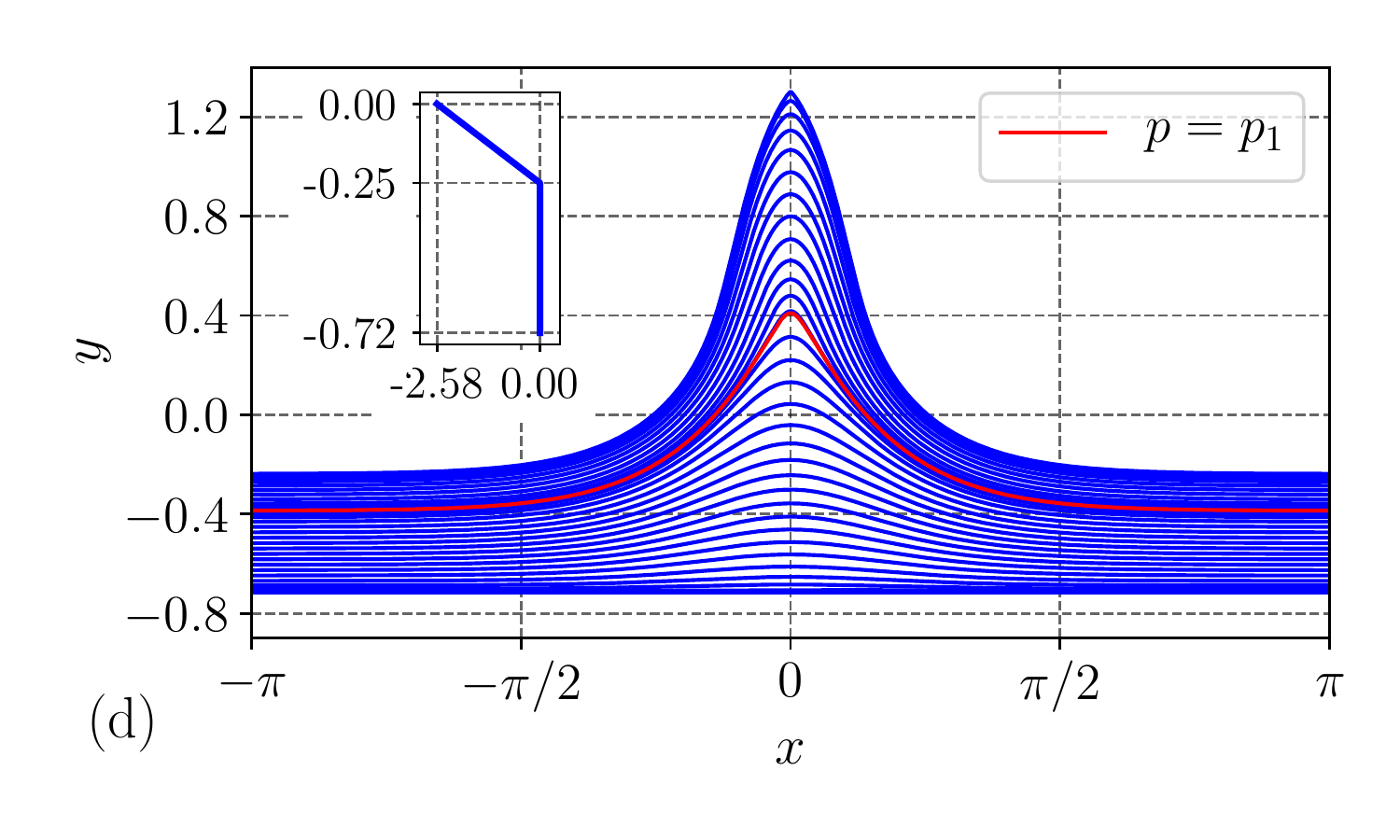} \\
   \end{array}$
   \caption{Streamlines of flows at the endpoints of each of the two parts of the bifurcation curve corresponding to $p_1=-0.7$, $\gm_1=-10.42$ (at the top layer), and $\gm_2=0$ (at the bottom layer): (a) laminar flow, (b) wave at the upper bound of the lower part, (c) wave at the lower bound of the upper part, and (d) wave at the upper bound of the upper part. In the figures, the insets show the background current profile when no wave motion.}
   \label{fig7:streamlines}
\end{figure}

\begin{figure}
   \centering$\begin{array}{cc}
   \includegraphics[trim={0.6cm 0.8cm 0 0},clip,height=3.8cm]{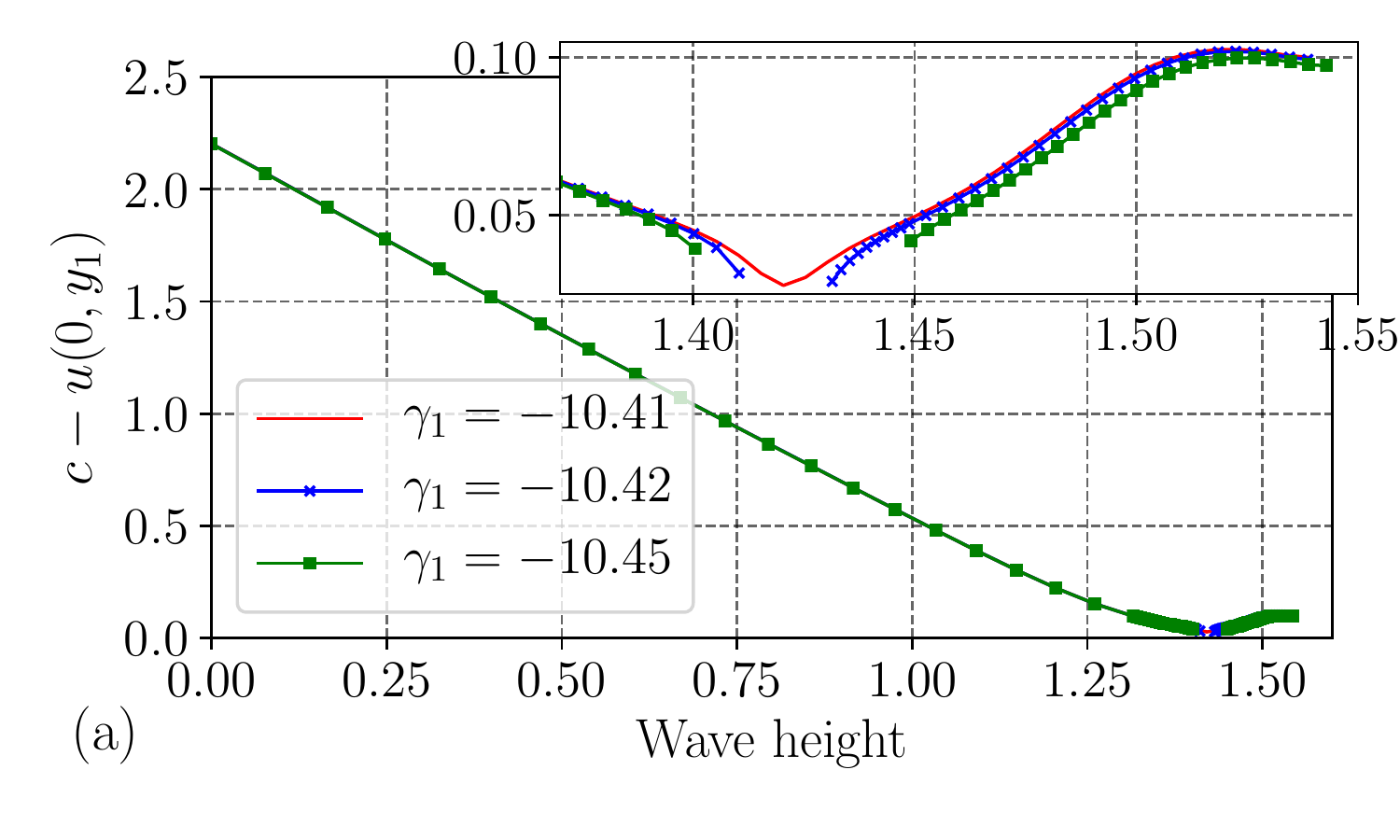} &
   \includegraphics[trim={0.6cm 0.8cm 0 0},clip,height=3.8cm]{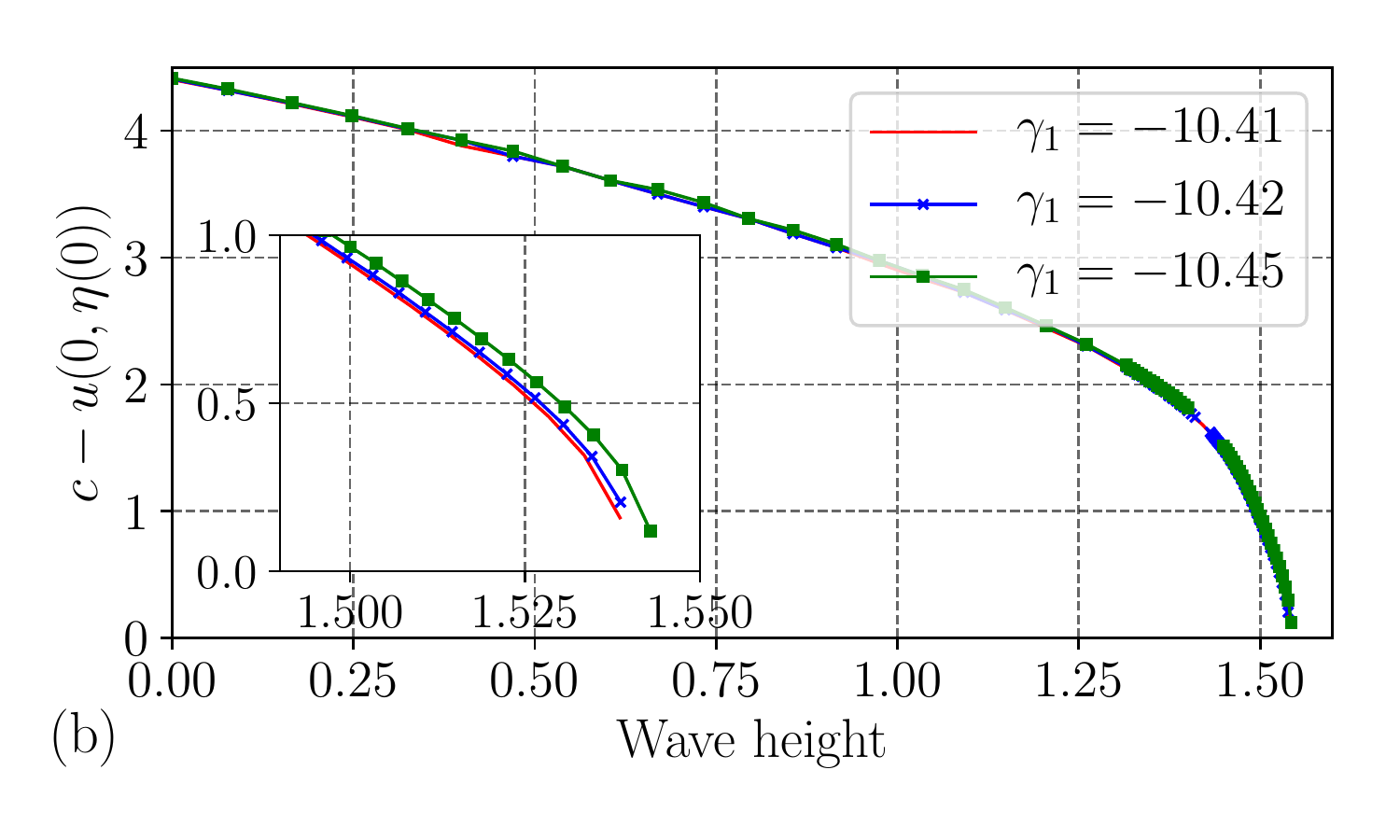} \\
   \end{array}$
   \caption{Horizontal velocity, $c-u$, under the crest along the bifurcation curve for $\gm_2=0$ (at the bottom layer), $p_1=-0.7$, and varying $\gm_1$ (at the top layer): (a) at the position where vorticity jumps ($y=y_1$ corresponds to $p=p_1$) and (b) at the surface. In the figures, the insets are closeup view of the curves when approaching the largest wave height.}
   \label{fig8:velocity}
\end{figure}

We further consider a layer ($p_1=-0.7$) of rotational flow with vorticity $\gm_1$ on an irrotational flow ($\gm_2=0$), following the setting in \cite{ko2008effect} and \cite{constantin2011periodic}. We found that the critical vorticity $\gm_{1,crit}$ is slightly below $-10.41$ (a value refined from $-10$ in \cite{ko2008effect}). By using $\gm_1$ as the bifurcation parameter, we also found the upper part of the branch of the bifurcation diagram when $\gm_1$ is less but very close to $\gm_{1,crit}$. Figure \ref{fig6:bifurcation} shows the bifurcation diagrams for $\gm_1=-10.41, -10.42$, and $-10.45$. For $\gm_1=-10.42$, the streamlines of the waves corresponding to the four endpoints of the bifurcation curve are shown in figure \ref{fig7:streamlines}, which clearly illustrates that the stagnation point first occurs internally and later occurs at the crest with further increased wave height. Figure \ref{fig8:velocity} shows the variation of the velocities $c-u$ at $p=p_1$ and at the surface along the bifurcation curves. The velocity $c-u$ at the position where vorticity jumps, is first decreasing and then increasing with respect to the wave height. Notably, the velocity decreases slightly again when close to the wave in flows with stagnation point at the surface. Again, the velocity $c-u$ on the crest is decreasing monotonically along with the increasing wave height. 

To summarize, in this subsection, we have observed new features of the bifurcation curves with fixed mass flux, for large amplitude waves on flows with discontinuous vorticity. Specifically, we show that along the bifurcation curve the stagnation point can first occur at the bottom or internally and then occur again at the surface for some particular vorticity distributions. 

\subsection{Pressure distribution}
We now present the pressure distribution for waves with stagnation points at the bottom, at an internal position and at the surface, for the case with discontinuous vorticity. 

\begin{figure}
   \centering
   \includegraphics[trim={0.2cm 0.2cm 0 0},clip,height=4cm]{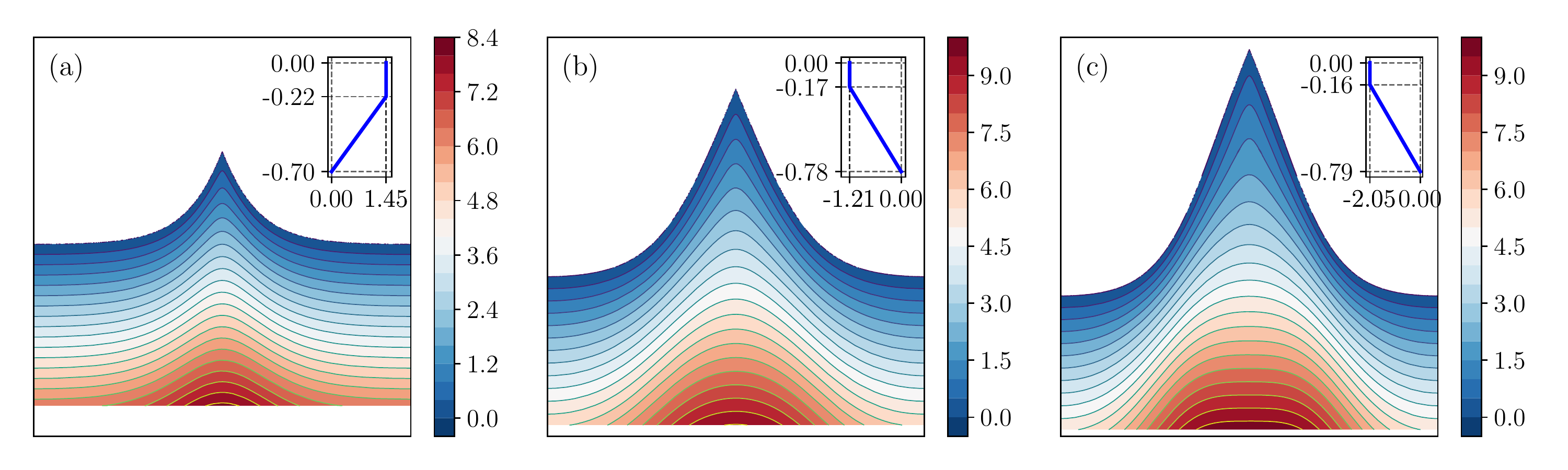}
   \caption{Pressure distribution for waves approaching waves with stagnation point at the crest for $p_1=-0.5$, $\gm_1=0$ (at the top layer), and: (a) $\gm_2=3$, (b) $\gm_2=-2$, and (c) $\gm_2=-3.23$. In the figure, the insets show the corresponding background shear flow when no wave motion.}
   \label{fig9:pressure}
\end{figure}

\begin{figure}
   \centering$\begin{array}{cc}
   \includegraphics[trim={0.5cm 0.2cm 0 0},clip,height=3.8cm]{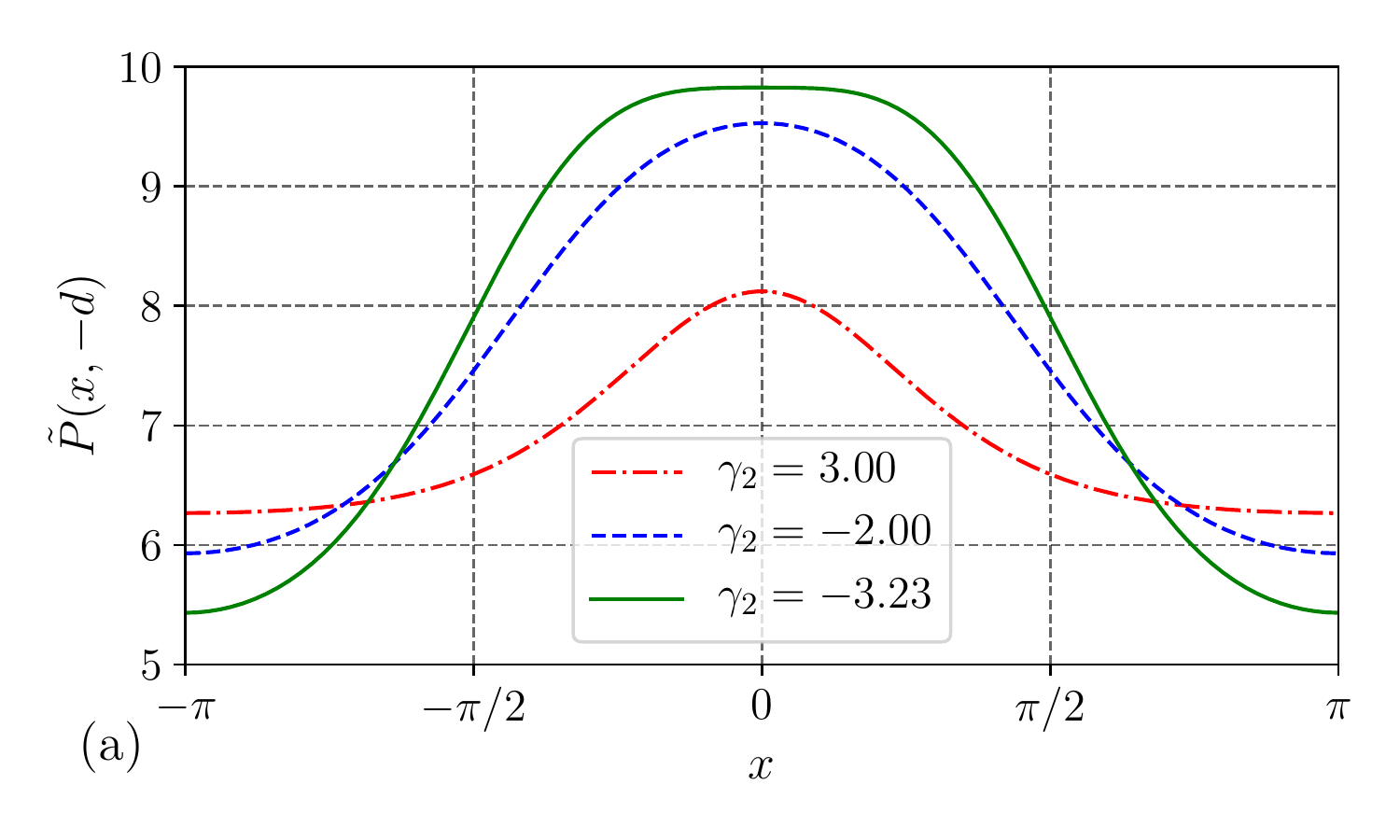} &
   \includegraphics[trim={0.5cm 0.2cm 0 0},clip,height=3.8cm]{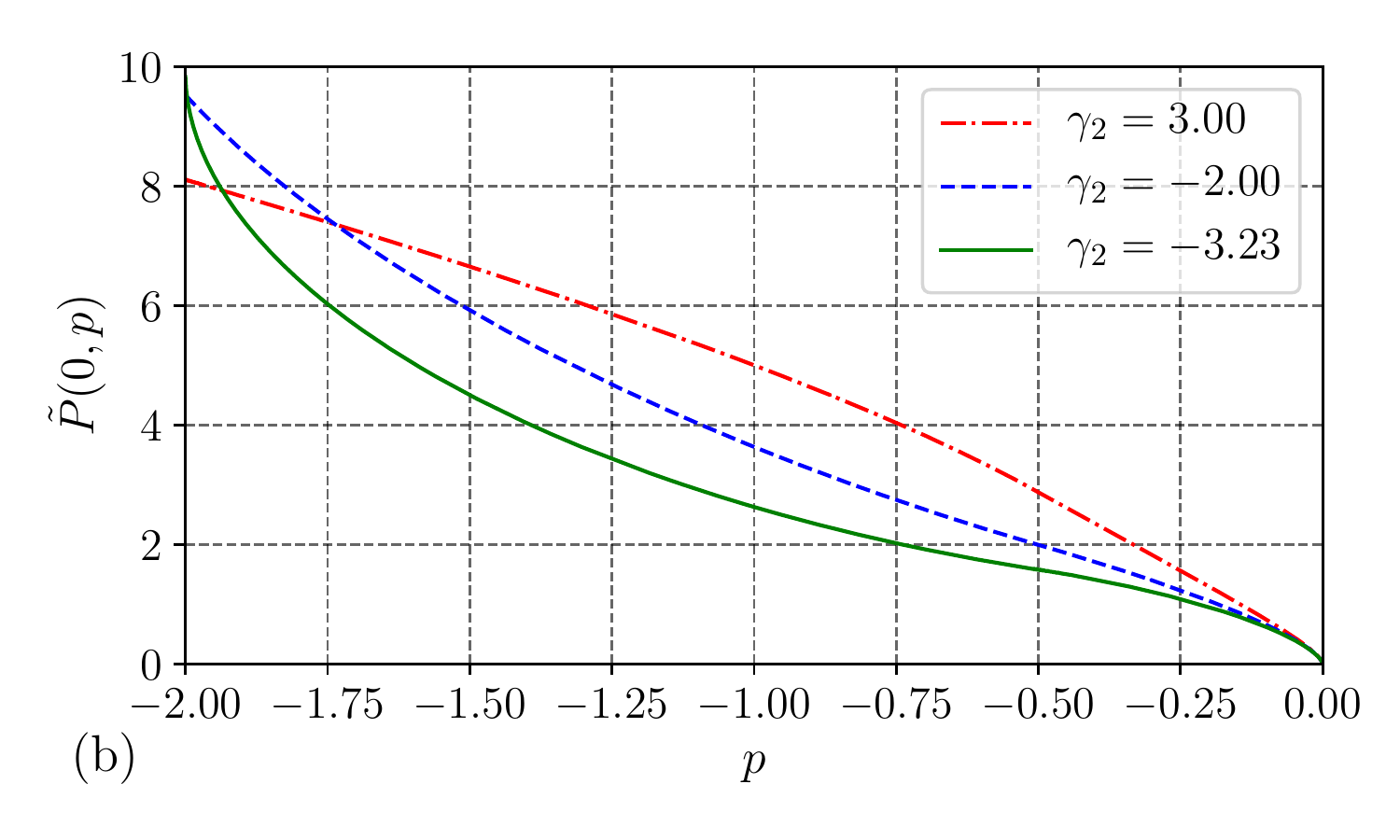} \\
   \end{array}$
   \caption{Pressure distribution in flows with an approaching stagnation point at the crest for $p_1=-0.5$, $\gm_1=0$ (at the top layer) and various different values of $\gm_2$ along (a) the bottom, and (b) the crest line.}
   \label{fig10:pressure}
\end{figure}

\begin{figure}
   \centering
   \includegraphics[trim={0.2cm 0.2cm 0 0},clip,height=4cm]{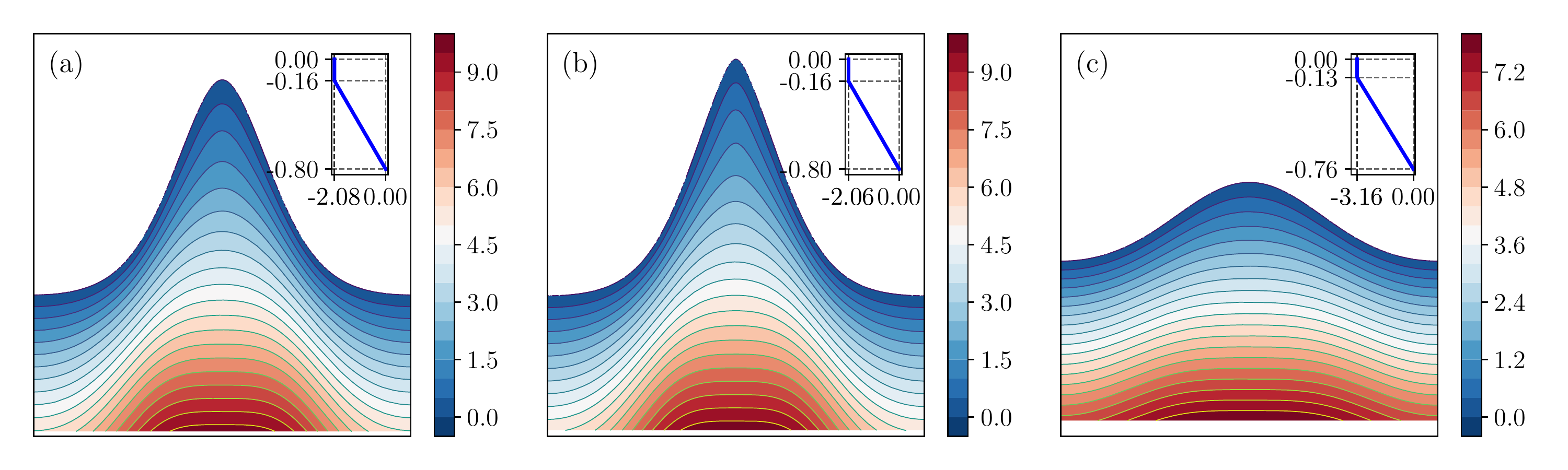}
   \caption{Pressure distribution  in flows with an approaching stagnation point at the bottom for $p_1=-0.5$, $\gm_1=0$ (at the top layer), and: (a) $\gm_2=-3.23$ (below gap), (b) $\gm_2=-3.23$ (above gap), and (c) $\gm_2=-5$. In the figure, the insets show the corresponding background shear flow when no wave motion.}
   \label{fig11:pressure}
\end{figure}

\begin{figure}
   \centering$\begin{array}{cc}
   \includegraphics[trim={0.5cm 0.2cm 0 0},clip,height=3.8cm]{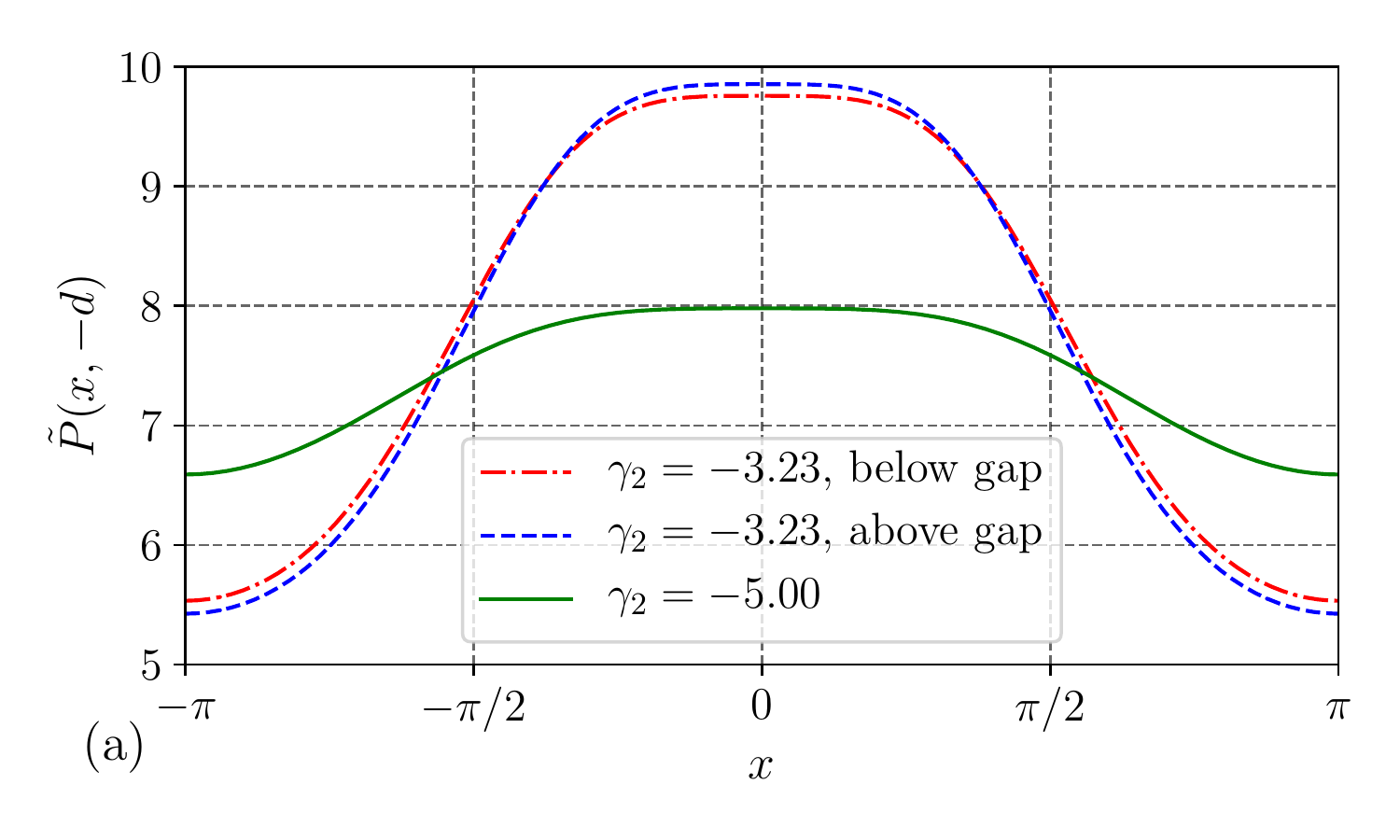} &
   \includegraphics[trim={0.5cm 0.2cm 0 0},clip,height=3.8cm]{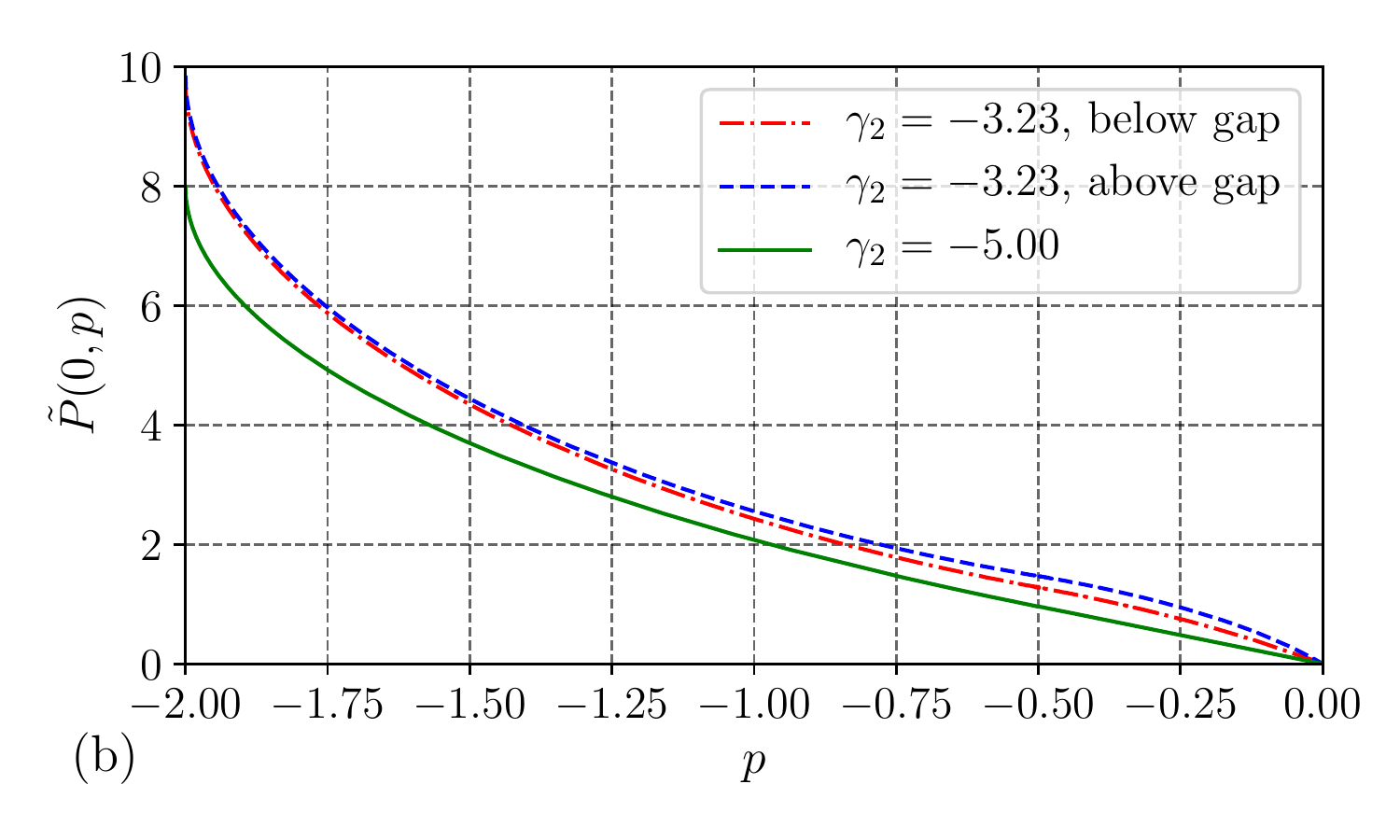} \\
   \end{array}$
   \caption{Pressure distribution  in flows with an approaching stagnation point  at the bottom for $p_1=-0.5$, $\gm_1=0$ (at the top layer) and various different values of $\gm_2$ along (a) the bottom, and (b) the crest line.}
   \label{fig12:pressure}
\end{figure}

\begin{figure}
   \centering
   \includegraphics[trim={0.2cm 0.2cm 0 0},clip,height=4cm]{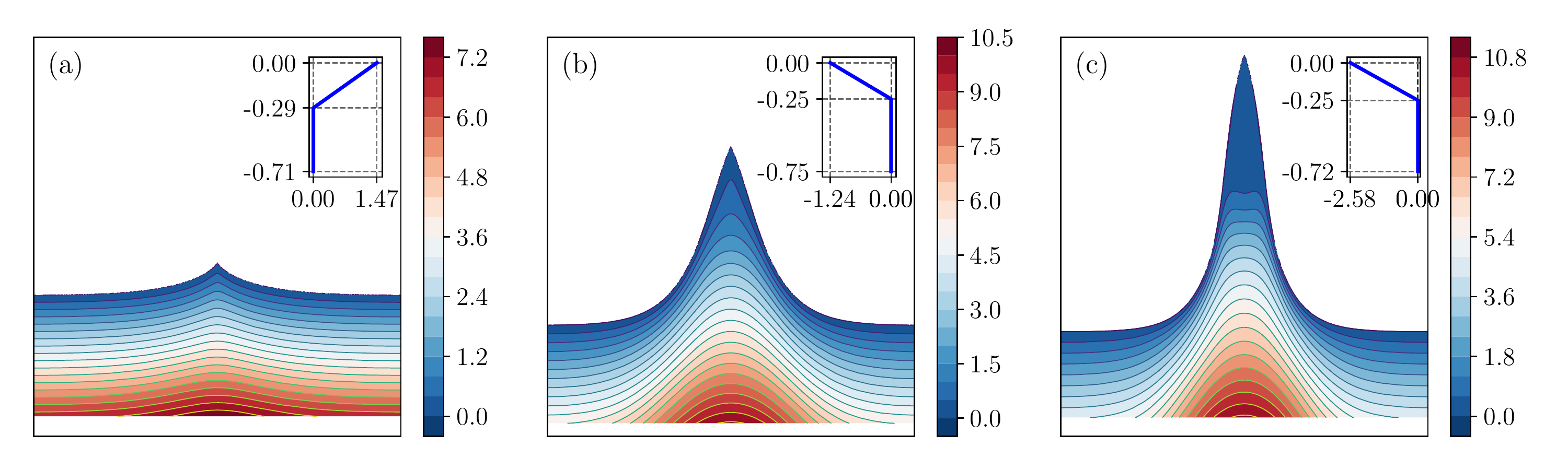}
   \caption{Pressure distribution  in flows with an approaching stagnation point  at the crest for $p_1=-0.7$, $\gm_2=0$ (at the bottom layer), and: (a) $\gm_1=5$, (b) $\gm_1=-5$, and (c) $\gm_1=-10.42$. In the figure, the insets show the corresponding background shear flow when no wave motion.}
   \label{fig13:pressure}
\end{figure}

\begin{figure}
   \centering$\begin{array}{cc}
   \includegraphics[trim={0.5cm 0.2cm 0 0},clip,height=3.8cm]{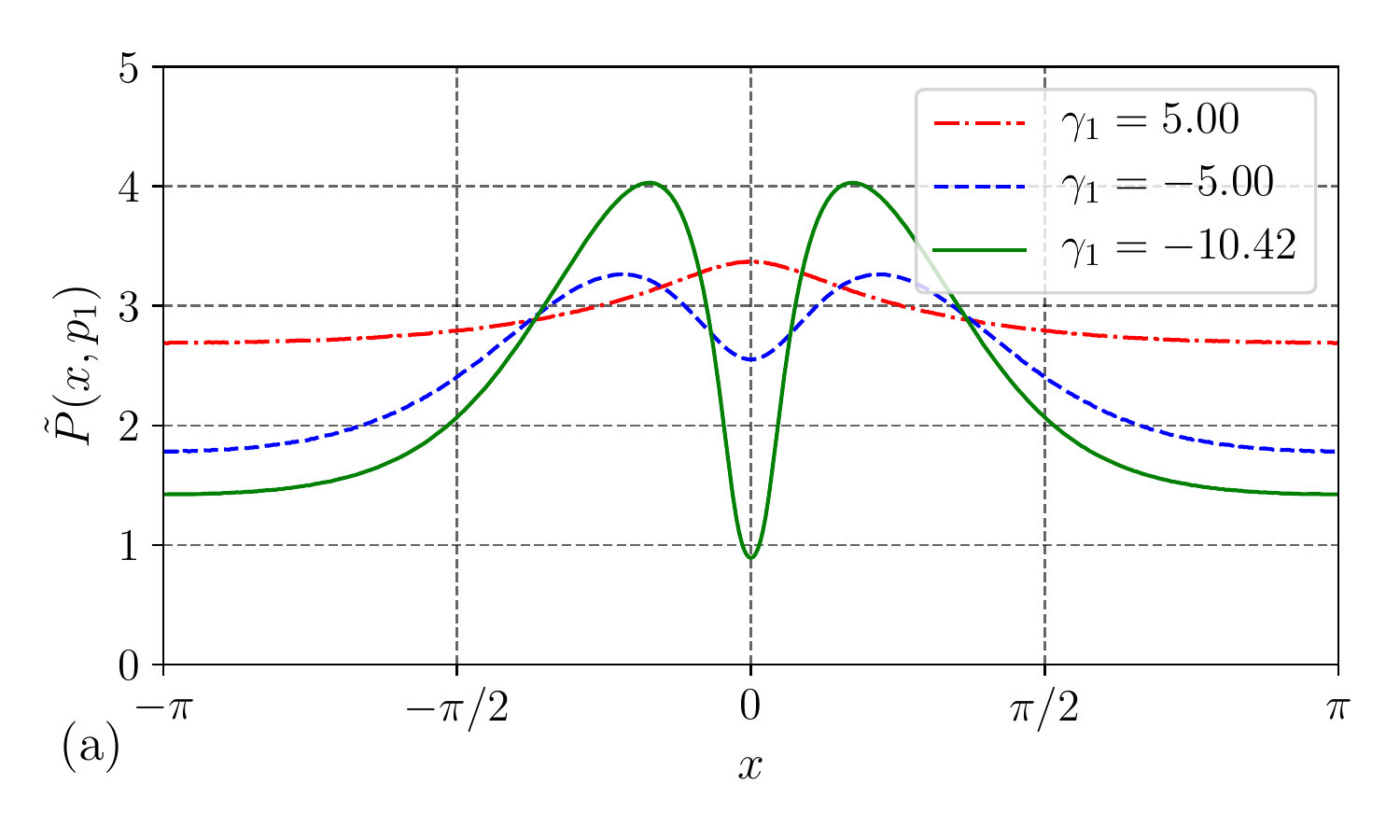} &
   \includegraphics[trim={0.5cm 0.2cm 0 0},clip,height=3.8cm]{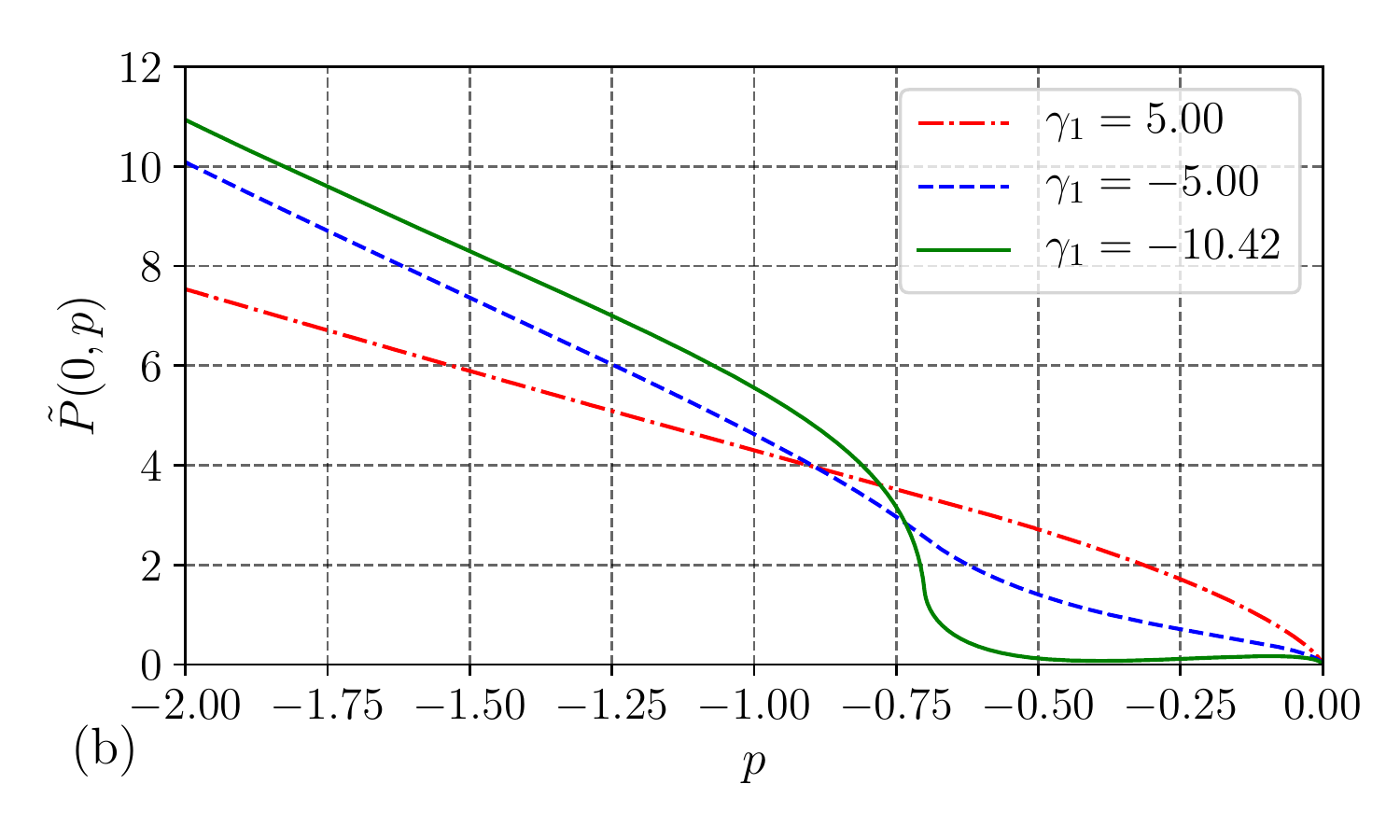} \\
   \end{array}$
   \caption{Pressure distribution in flows with an approaching stagnation point  at the crest for $p_1=-0.7$, $\gm_2=0$ (at the bottom layer) and various different values of $\gm_1$ along (a) the stream line $\psi=-p_1$, and (b) the crest line.}
   \label{fig14:pressure}
\end{figure}

\begin{figure}
   \centering
   \includegraphics[trim={0.2cm 0.2cm 0 0},clip,height=4cm]{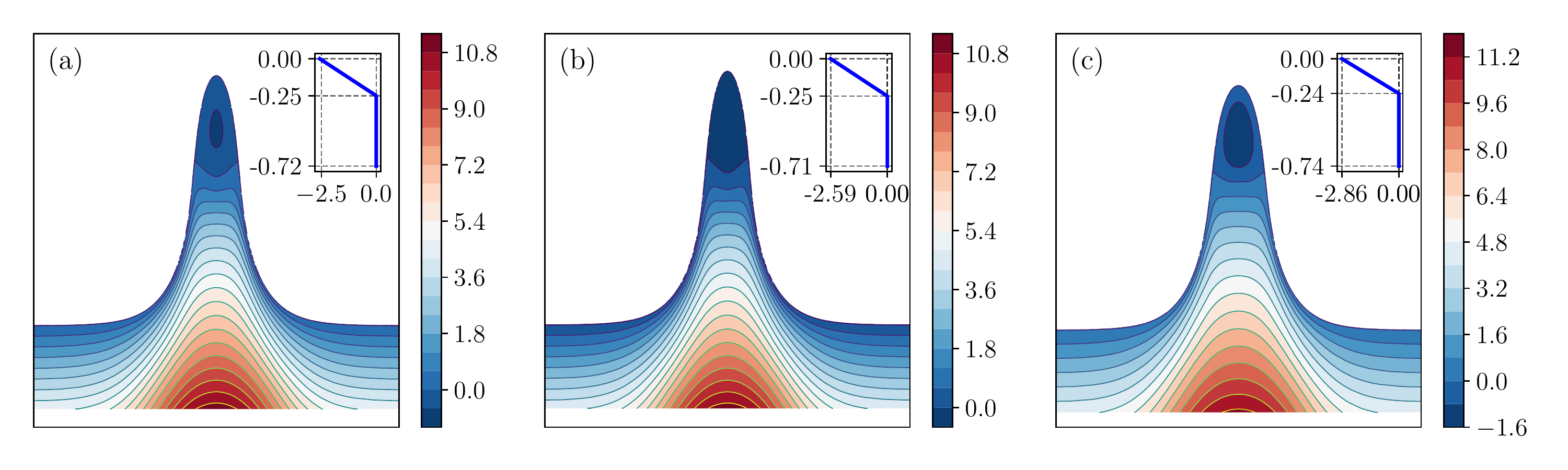}
   \caption{Pressure distribution in flows with an approaching stagnation point  at a position where vorticity jump occurs, for $p_1=-0.7$, $\gm_2=0$ (at the bottom layer), and: (a) $\gm_1=-10.42$ (below gap), (b) $\gm_1=-10.42$ (above gap), and (c) $\gm_1=-12$. In the figures, the insets show the background current profile when no wave motion.}
   \label{fig15:pressure}
\end{figure}

\begin{figure}
   \centering$\begin{array}{cc}
   \includegraphics[trim={0.5cm 0.2cm 0 0},clip,height=3.8cm]{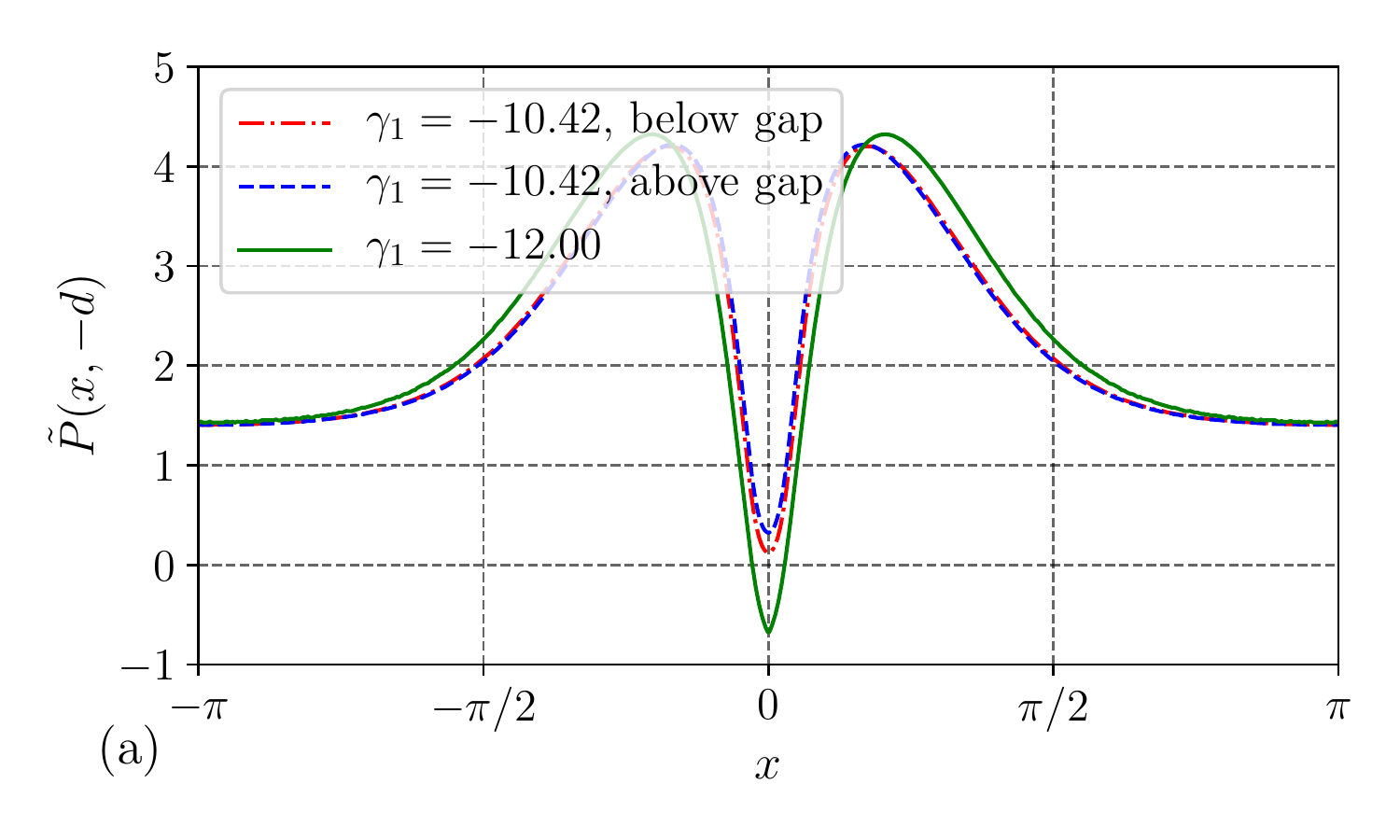} &
   \includegraphics[trim={0.5cm 0.2cm 0 0},clip,height=3.8cm]{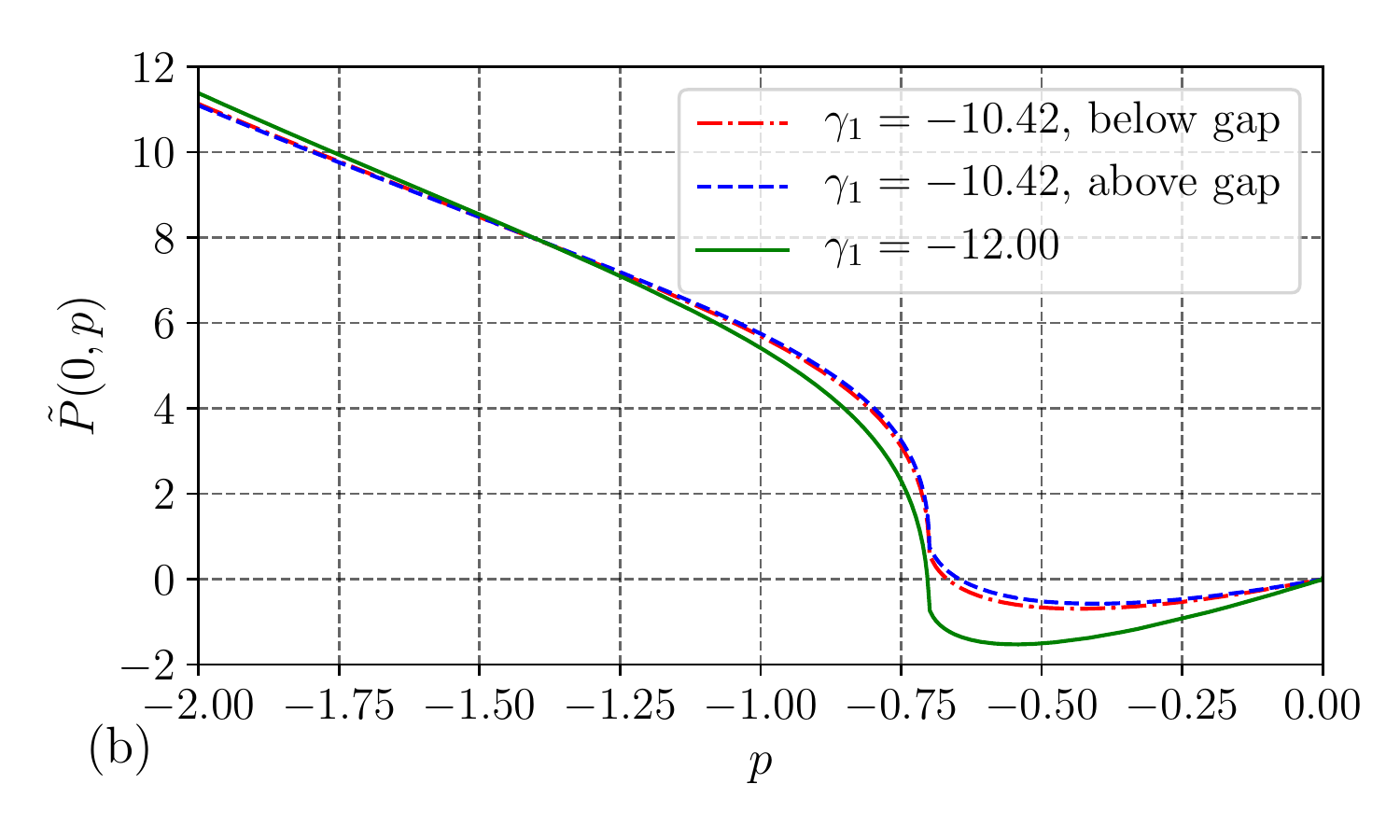} \\
   \end{array}$
   \caption{Pressure distribution in flows with an approaching stagnation point  at a position where vorticity jump occurs for $p_1=-0.7$, $\gm_2=0$ and various different values of $\gm_1$ along (a) the stream line $\psi=-p_1$, and (b) the crest line.}
   \label{fig16:pressure}
\end{figure}

For an irrotational flow on a rotational flow with $p_1=-0.5$ and $\gm_1=0$, we consider the cases corresponding to $\gm_2=3, -2$ and $-5$. For waves close to the case of flows 
having a stagnation point at the crest, the pressure distributions are shown in figure \ref{fig9:pressure}, and figure \ref{fig10:pressure} compares the pressure along the bottom and along the crest line. For waves in flows close to having a stagnation point at the bottom, the pressure distribution demonstrated in figure \ref{fig11:pressure} and figure \ref{fig12:pressure} shows the pressure along the bottom and the crest line. Two features of the pressure distributions are consistent with the pressure beneath a Stokes wave as proved by \cite{constantin2010pressure} and those can be stated as \lq{}\lq{the pressure in the fluid strictly decreases horizontally away from a crest line toward its neighboring trough lines, and the pressure strictly increases with depth}\rq{}\rq{}. Even though, when $\gm_2$ is close to the critical value, the pressure along the bottom displays a \lq\lq{plateau}\rq\rq{} behavior, as previously pointed out by \cite{amann2018numerical} for the case of constant vorticity, the pressure is still decreasing away from the crest line, slowly.

\begin{figure}
    \centering$\begin{array}{cc}
   \includegraphics[trim={0.5cm 0.2cm 0 0},clip,height=3.8cm]{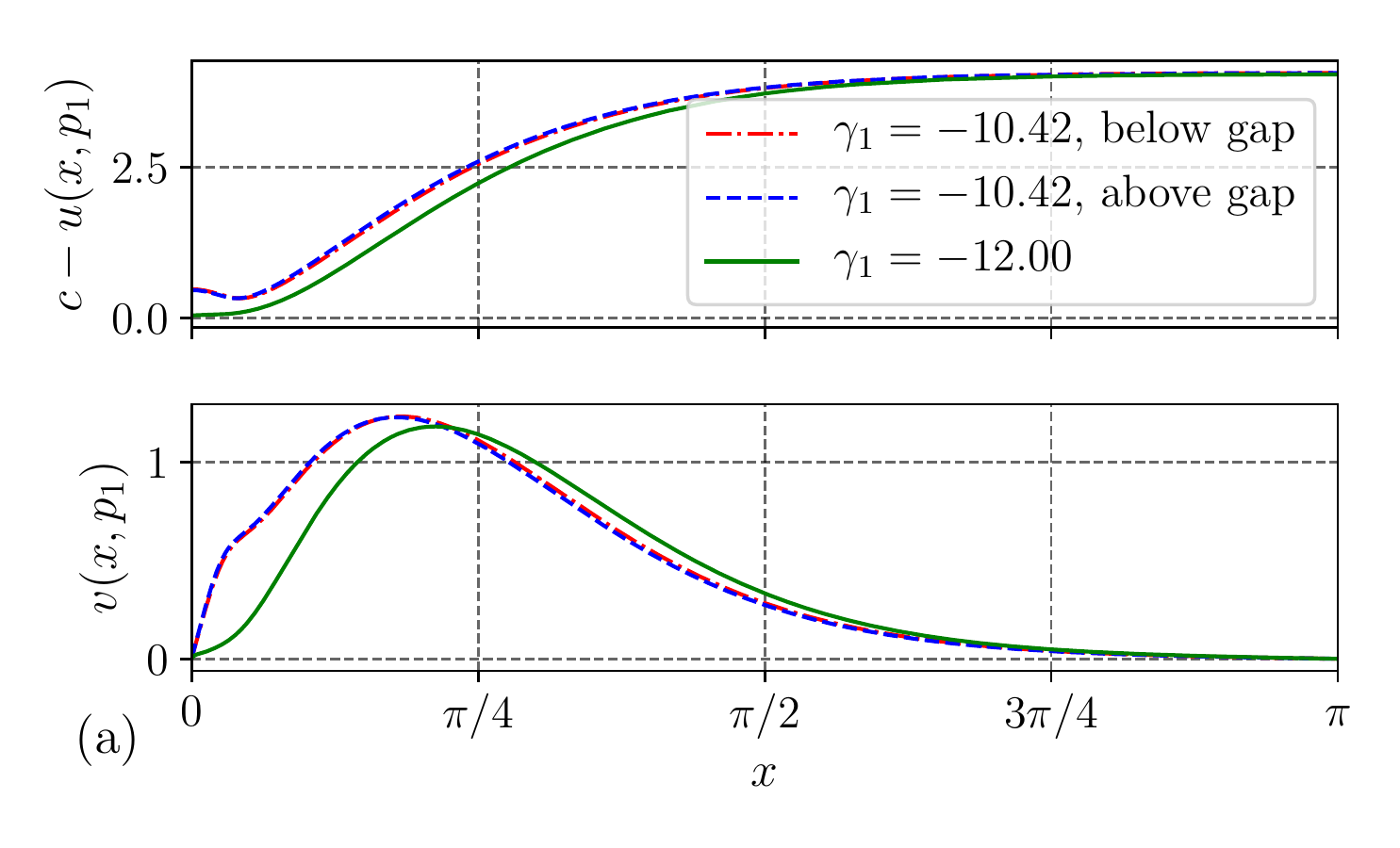} &
   \includegraphics[trim={0.5cm 0.2cm 0 0},clip,height=3.8cm]{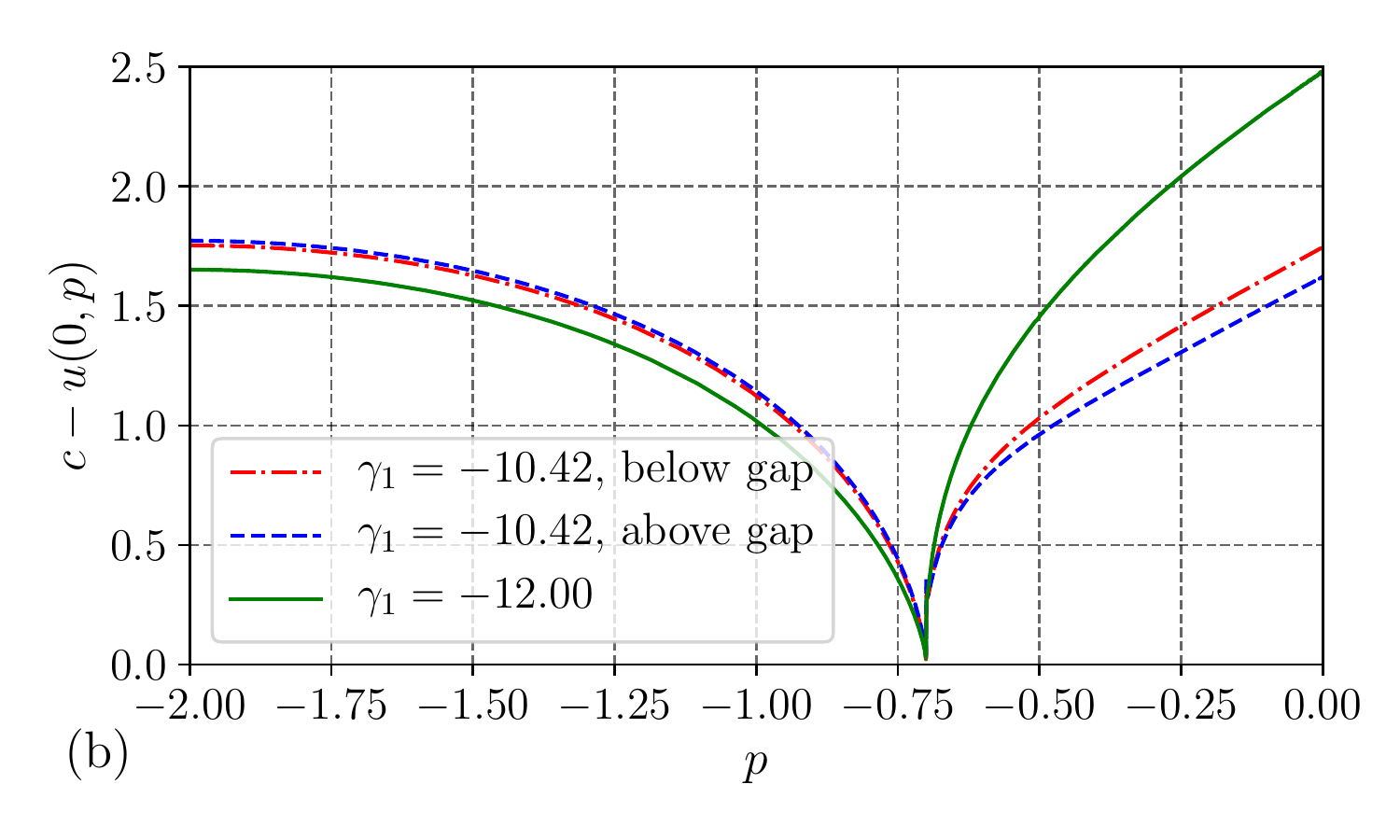} \\
   \end{array}$
   \caption{Variation of velocities, $c-u$ and $v$, with  $p_1=-0.7$, $\gm_2=0$ (at the bottom layer), and various different values of $\gm_1$: (a) $c-u$ and $v$ along the streamline $\psi=-p_1$, and (b) $c-u$ along the crest line ($v$ is zero along this line and hence not shown).}
   \label{fig17:velocity}
\end{figure}

For a rotational flow on an irrotational flow, we analyze the cases for $\gm_1=5, -5$, and $-10.42$. Figure \ref{fig13:pressure} shows the pressure distribution in flows close to having a stagnation point at the crest, and variations of the pressure along the bottom and along the crest line are illustrated in figure \ref{fig14:pressure}. The pressure distribution in flows when internal stagnation points are almost developed is illustrated in figure \ref{fig15:pressure}. Correspondingly, figure \ref{fig16:pressure} shows the pressure along the streamline $\psi=-p_1$ and along the crest line. From these figures, it is seen that none of the facts regarding the pressure distribution for an irrotational Stokes wave \citep{constantin2010pressure} hold, as also observed by \cite{ribeiro2017flow} for rotational flows with constant vorticity. When the vorticity $\gm_1$ is below a critical value, the pressure along the streamline $\psi=-p_1$ achieves a local minimum under the crest, as opposed to the maximum there when $\gm_1$ is large. For smaller $\gm_1$, the relative pressure $\tilde P$ at this position becomes negative (i.e. sub-atmospheric), which is the minimum along the streamline. The pressure attaining a value less than the atmospheric pressure has also been observed by \cite{ali2013reconstruction} for the case of constant vorticity using a model based on an asymptotic expansion for the stream functions. When $\gm_1$ is below a critical value, the pressure below but close to the layer where the vorticity jump occurs also exhibits the \lq{}\lq{plateau}\rq{}\rq{} behavior, internally, see figure \ref{fig15:pressure}. Figures \ref{fig14:pressure}-\ref{fig16:pressure} show that when $\gm_1$ is much larger than the critical value, the pressure is increasing with depth along the crest line. Interestingly, when $\gm_1$ is quite close to $\gm_{1,crit}$, the pressure $\tilde P$ along the crest line first increases slightly, and then decreases, and increases again when close to the position where vorticity jump occurs; the overall pressure in the surface layer is close to the atmospheric pressure, see figure \ref{fig14:pressure}b for $\gm_1=-10.42$. With a further decrease in the value of $\gm_1$ below the critical value, the relative pressure in the surface layer becomes negative, first decreasing from the crest and then increasing near the position where the vorticity jump takes place. To understand the reason for the development of a negative relative pressure $\tilde P$ in the surface layer near the crest line, we recall the expression \eqref{eq:P}.  For the cases illustrated in figure \ref{fig14:pressure}, the term $\int_0^p\gm(-s)\df s$ in \eqref{eq:P} is positive in the surface layer and increases rapidly with depth since $|\gm_1|$ is large, the term $\grav(y+d)$ is also large considering the large wave height, and the velocity $c-u$ is also not small, as shown in figure \ref{fig17:velocity}; hence $\tilde P$ in the surface layer becomes negative. Further decreasing $\gm_1$ (with increasing $|\gm_1|$), it seems that the minimal pressure along the crest line will keep decreasing. 


\subsection{Particle trajectories}
We proceed to compute the particle trajectories under the waves corresponding to flows with nearly developed stagnation points, for the previously mentioned cases. To recover the wave speed using expression \eqref{eq:speed}, we assume that the current strength at the bed is zero, i.e., $\kappa=0$. Note that the choice of a different setting (adding an arbitrary uniform current) may lead to variations of the resulting particle paths. For example, when studying particle paths under waves on linear shear flow, \cite{ribeiro2017flow} set the current strength at the mid-depth to be zero (to ensure a flow with no net mass flux in the case of zero-amplitude waves). However, we here focus on the relative horizontal displacements of the particles along the depth, which are not affected by the presence of a uniform background current.

\begin{figure}
   \centering
   \includegraphics[trim={0.2cm 0.2cm 0 0},clip,height=4cm]{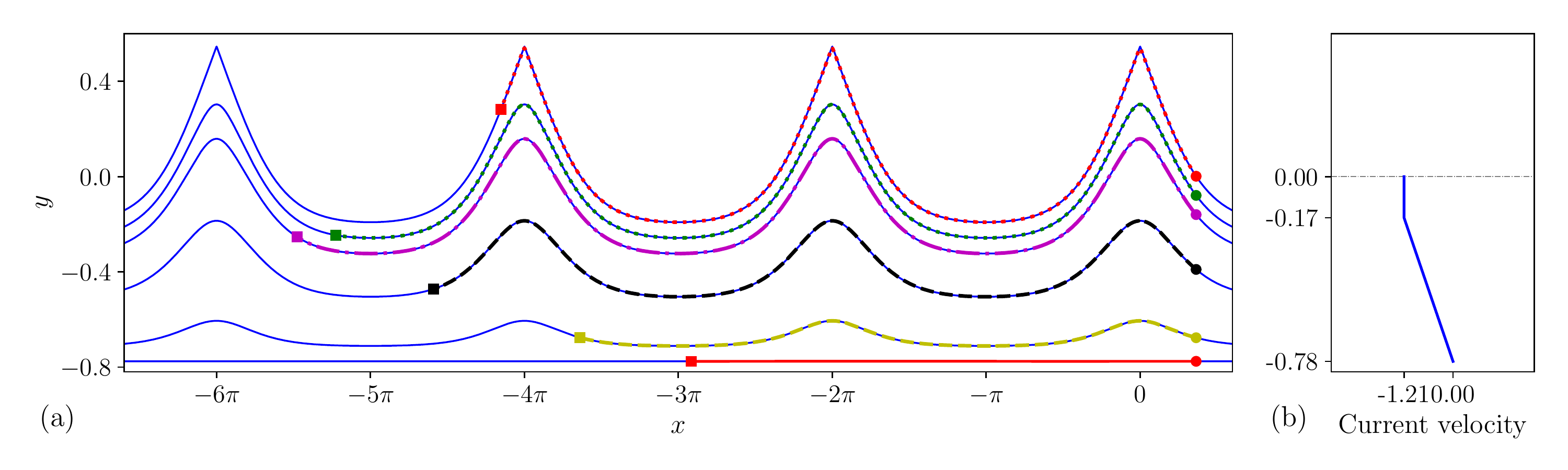}
   \caption{(a) Particle trajectories under the waves in flows close to having a stagnation point at the crest with $p_1=-0.5$, $\gm_1=0$, and $\gm_2=-2$ in the moving frame; (b) the background current profile when no wave motion.}
   \label{fig18:path}
\end{figure}

In the following computation and illustration of the particle trajectories, we set that: (i) the initial positions of the particles are under a zero-crossing point of the wave profile \citep{paprota2018particle}; (ii) the particles trajectories are plotted for the waves traveling rightwards for two wave periods, i.e., for a time of $4L/c=4\pi/c$, and the final position of 
the particles under zero-amplitude waves are also indicated in the graphs for comparison. In the graphs, six particle trajectories are plotted. Three of those particles are located at the surface, at the interface, and at a position very close to the bottom but not exactly at the bottom. The other three particles include one located in the top layer and two located in the bottom layer. The initial positions are denoted by dots and the final positions are indicated by square markers. 

\begin{figure}
   \centering 
   $\begin{array}{c}
   \includegraphics[trim={0.4cm 0.2cm 0 0},clip,height=4cm]{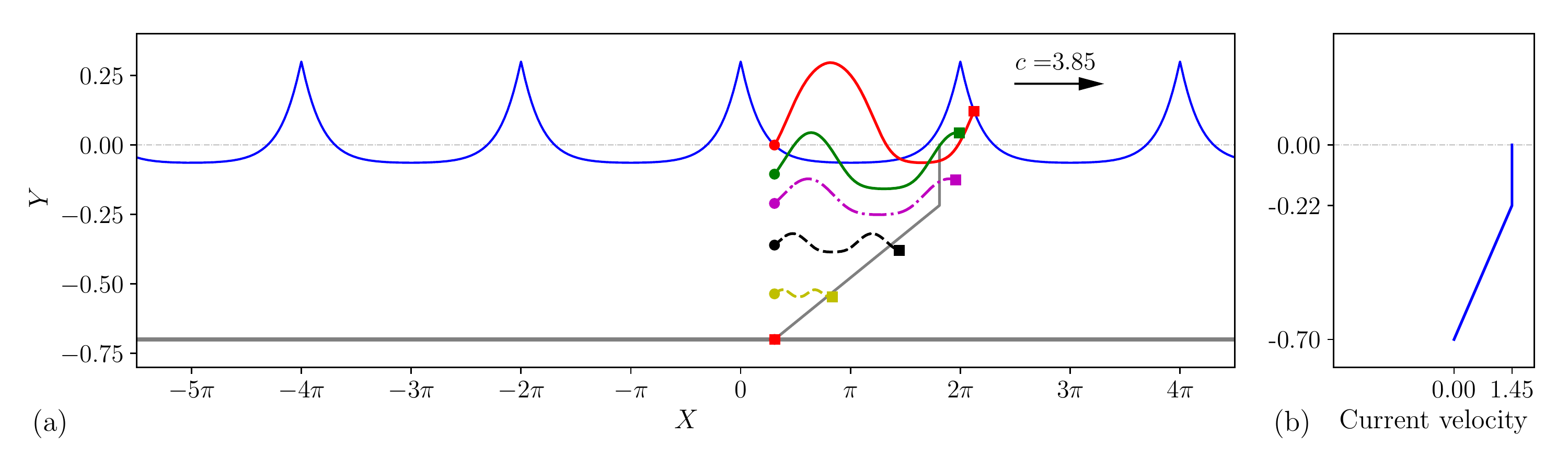}\\
   \includegraphics[trim={0.4cm 0.2cm 0 0},clip,height=4cm]{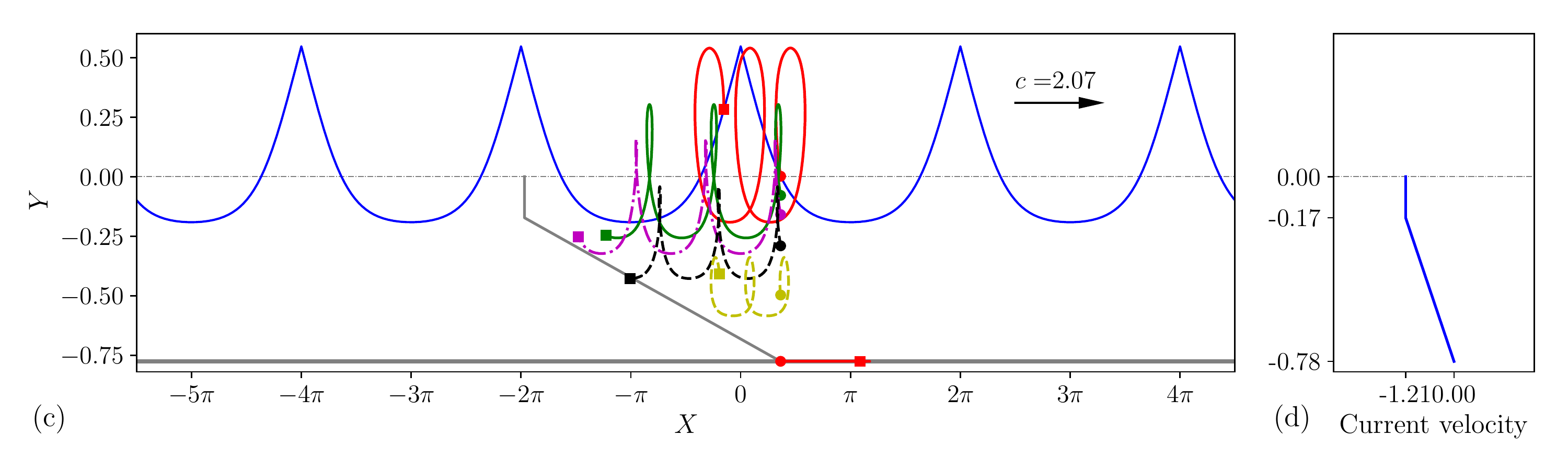}\\
   \includegraphics[trim={0.4cm 0.2cm 0 0},clip,height=4cm]{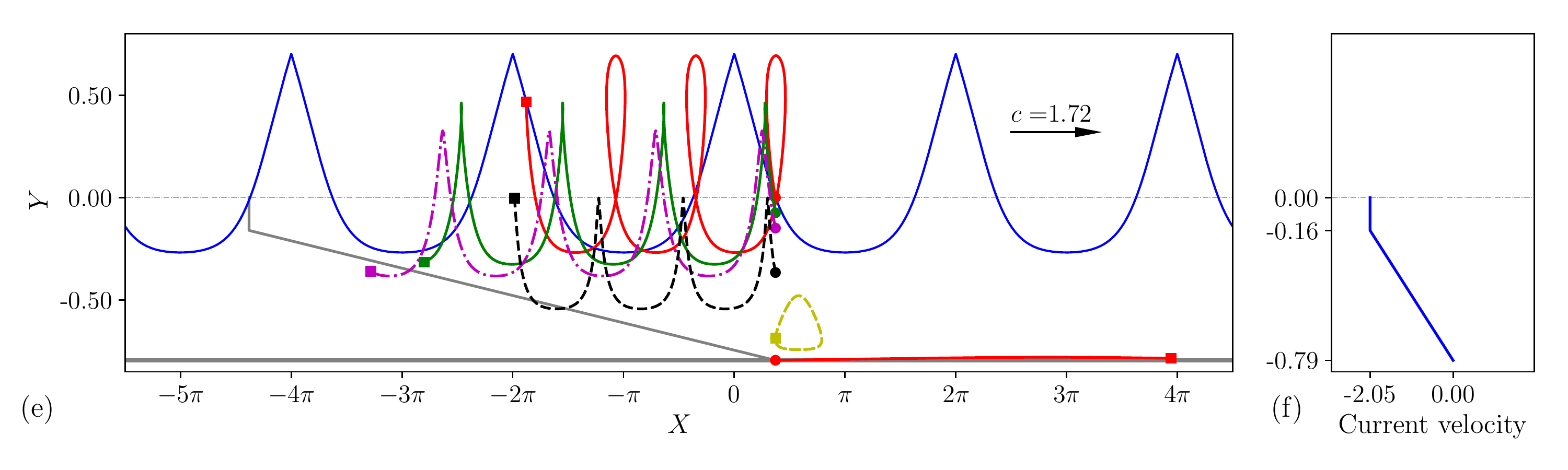}\\
   \end{array}$
   \caption{Particle trajectories under the waves in flows close to having a stagnation point near the crest, for $p_1=-0.5$ and $\gm_1=0$: (a) $\gm_2=3$; (c) $\gm_2=-2$; (e) $\gm_2=-3.23$. The background current profiles corresponding to zero-amplitude waves are shown in (b), (d), and (f), respectively. In the plot of the particle paths, the gray lines indicate the final position of the particles if under zero-amplitude waves after the time $4L/c$.}
   \label{fig19:paths}
\end{figure}

\begin{figure}
   \centering 
   $\begin{array}{c}
   \includegraphics[trim={0.4cm 0.2cm 0 0},clip,height=4cm]{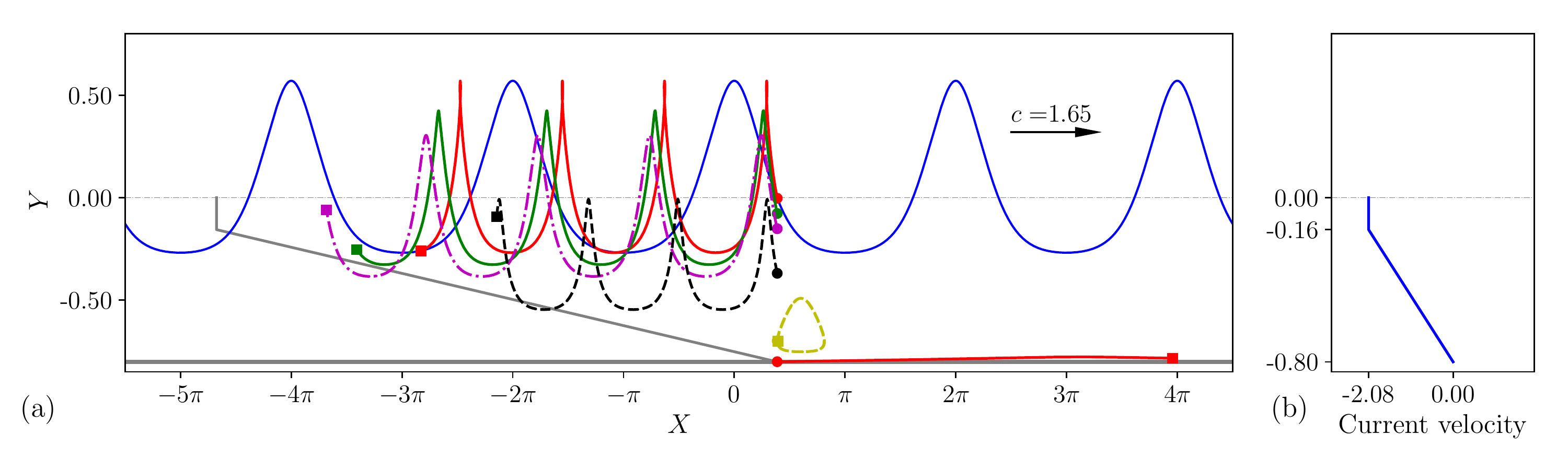}\\
   \includegraphics[trim={0.4cm 0.2cm 0 0},clip,height=4cm]{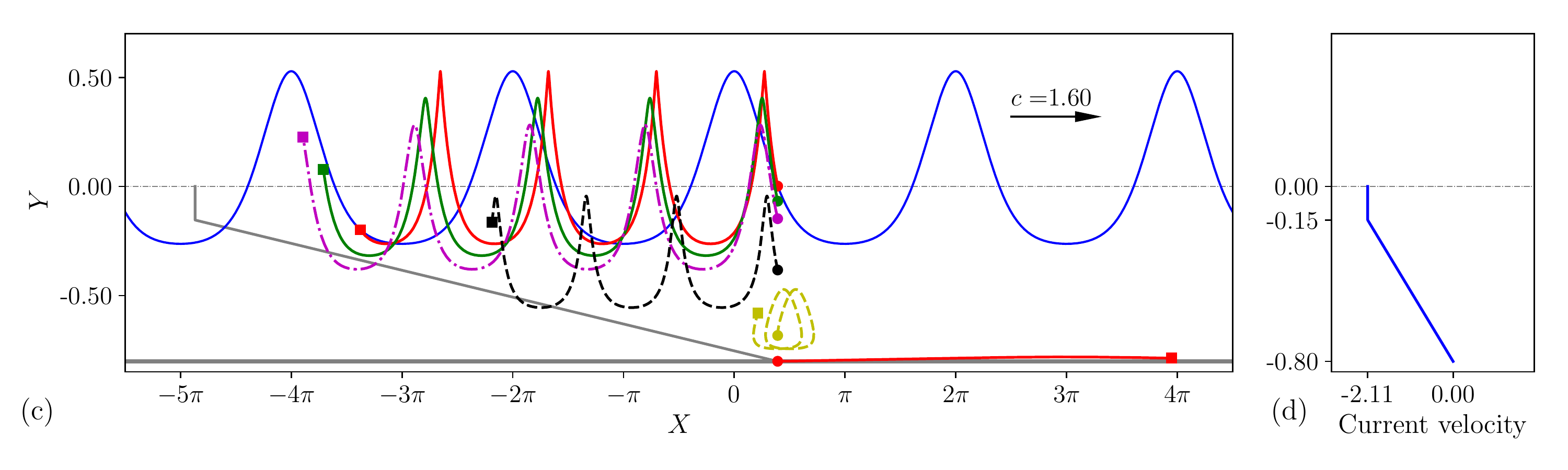}\\
   \end{array}$
   \caption{Particle trajectories under the waves in flows close to having a stagnation point at the bottom, for $p_1=-0.5$ and $\gm_1=0$: (a) $\gm_2=-3.23$ (below the gap); (c) $\gm_2=-3.25$ (below the gap). The background current profiles corresponding to zero-amplitude waves are shown in (b) and (d), respectively. In the plot of the particle paths, the gray lines indicate the final position of the particles if under zero-amplitude waves after the time $4L/c$.}
   \label{fig20:paths}
\end{figure}

\begin{figure}
   \centering 
   $\begin{array}{c}
   \includegraphics[trim={0.4cm 0.2cm 0 0},clip,height=4cm]{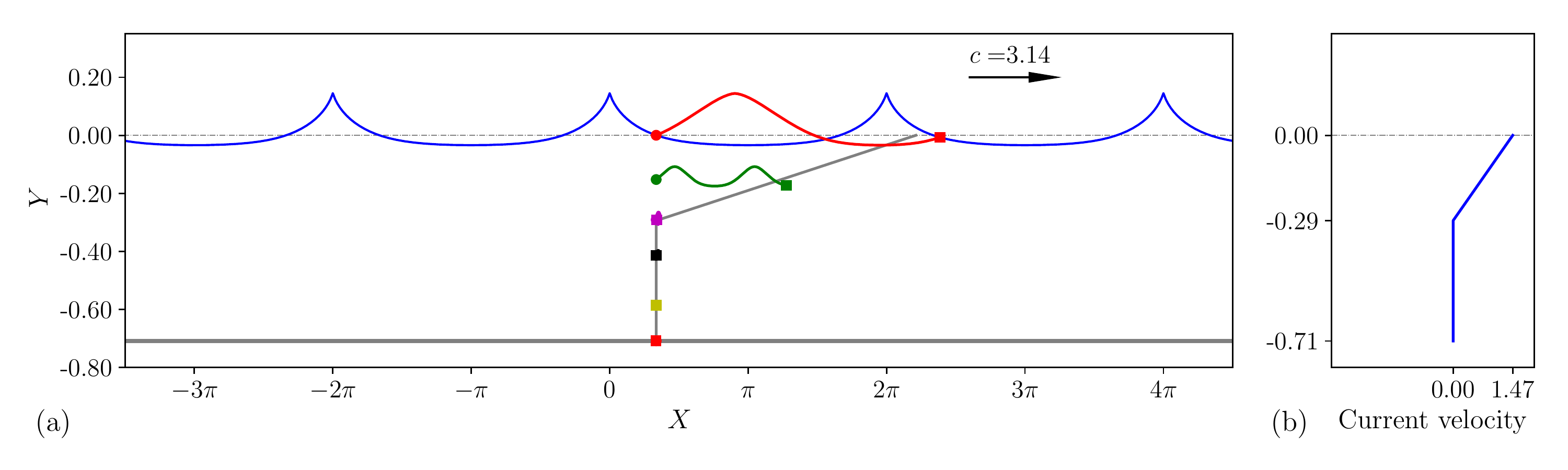}\\
   \includegraphics[trim={0.4cm 0.2cm 0 0},clip,height=4cm]{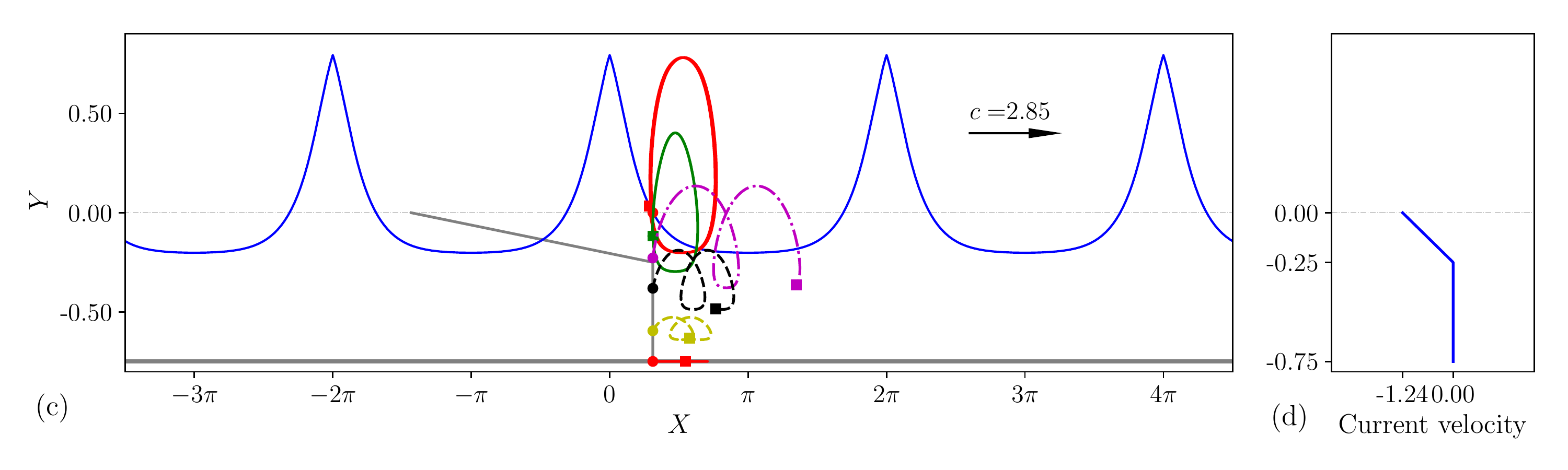}\\
   \includegraphics[trim={0.4cm 0.2cm 0 0},clip,height=4cm]{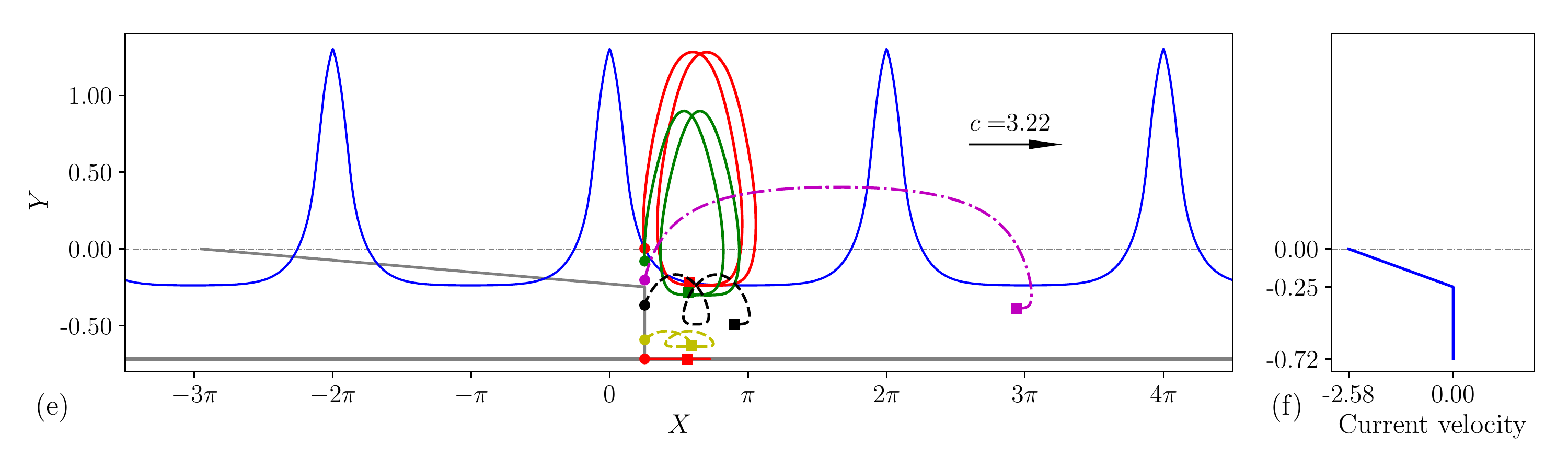}\\
   \end{array}$
   \caption{Particle trajectories under the waves in flows close to having a stagnation point near the crest, for $p_1=-0.7$ and $\gm_2=0$: (a) $\gm_1=5$; (c) $\gm_1=-5$; (e) $\gm_1=-10.42$. The background current profiles corresponding to zero-amplitude waves are shown in (b,d,f), respectively. In the plot of the particle paths, the gray lines indicate the final position of the particles if under zero-amplitude waves after the time $4L/c$.}
   \label{fig21:paths}
\end{figure}

\begin{figure}
   \centering 
   $\begin{array}{c}
   \includegraphics[trim={0.4cm 0.2cm 0 0},clip,height=4cm]{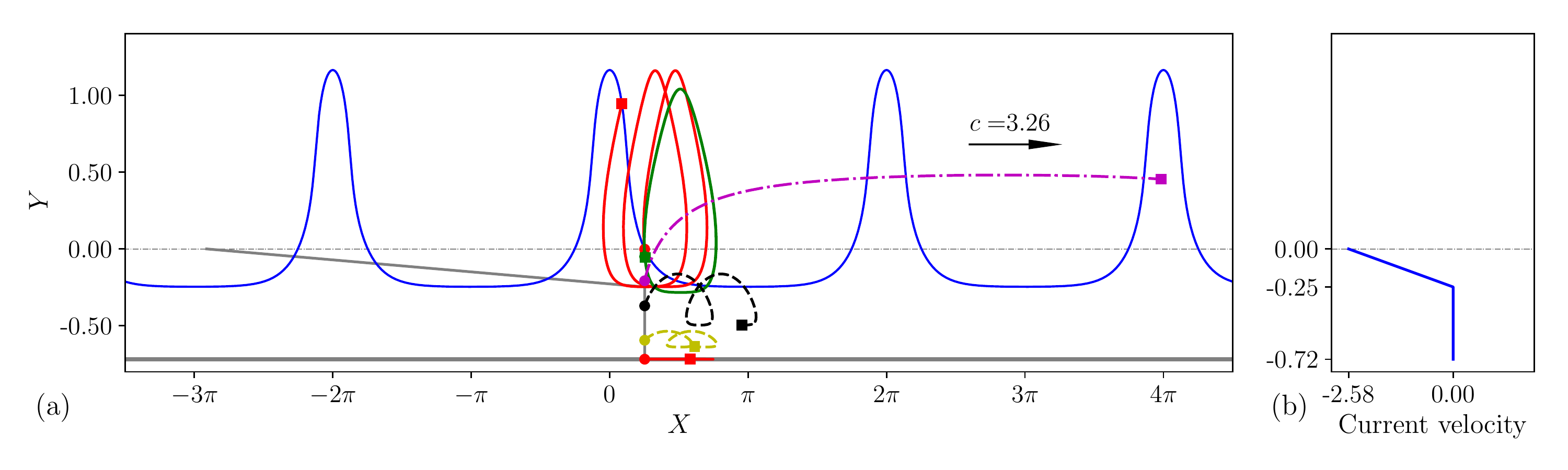}\\
   \includegraphics[trim={0.4cm 0.2cm 0 0},clip,height=4cm]{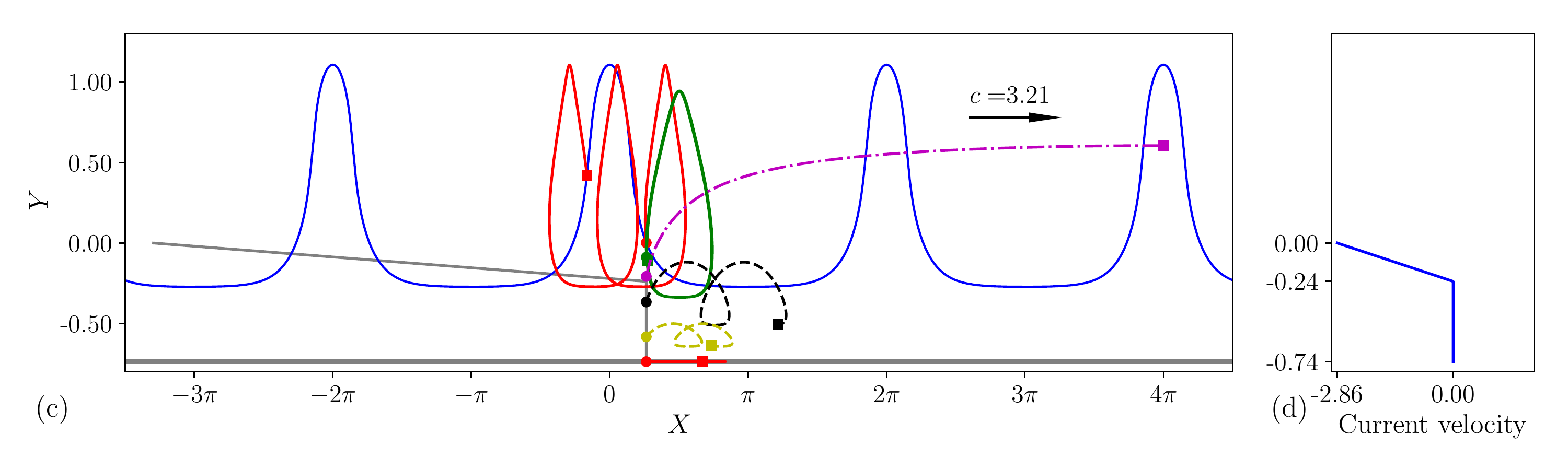}\\
   \end{array}$
   \caption{Particle trajectories under the waves in flows close to having a stagnation point at the bottom, for $p_1=-0.7$ and $\gm_2=0$: (a) $\gm_2=-10.42$ (below the gap); (c) $\gm_2=-12$. The background current profiles corresponding to zero-amplitude waves are shown in (b) and (d), respectively. In the plot of the particle paths, the gray lines indicate the final position of the particles if under zero-amplitude waves after the time $4L/c$.}
   \label{fig22:paths}
\end{figure}

First of all, we plot the computed particle trajectories in the moving frame, as illustrated in figure \ref{fig18:path}, which theoretically are the corresponding streamlines. The streamlines passing through the initial particle locations are plotted for comparison, and a good agreement is seen, suggesting the accuracy of the method for computing particle trajectories.

For an irrotational layer on a rotational layer, the particle trajectories in flows close to having a stagnation point at the crest are plotted in figure \ref{fig19:paths} and those in flows close to having a stagnation point at the bottom are plotted in figure \ref{fig20:paths}. We show that as theoretically predicted by \cite{constantin2006trajectories} when a particle reaches 
the crest of the wave in flows having a stagnation point at the crest, it does not pause there but moves further (figure \ref{fig19:paths}a). However, for waves in flows close to having 
an internal stagnation point or a stagnation point at the bottom, a particle reaching the stagnation point tends to pause at that point in the moving frame, similar to the case shown in \cite{ribeiro2017flow} for flows 
with a bottom stagnation point. Some observations on the features of the particle trajectories are listed below:
\begin{enumerate}
\item In both figure \ref{fig19:paths} and figure \ref{fig20:paths}, the horizontal position of the particles are on the right of the gray lines (which are the position of the particles with no waves), indicating that the waves induce a positive drift of the particles in the direction of the propagating waves. This is consistent with what was proved by \cite{constantin2012particle} and the experimental results \citep{paprota2018particle} for irrotational flows. 
\item In the top irrotational layer, the horizontal displacement (with respect to the background current induced displacement) or dirft is larger when the particle is closer to the surface than when it is near the interface of the two layers, which is consistent with experimental and approximate theoretical results \citep{paprota2018particle,curtis2018particle}. In the bottom rotational layer, when the vorticity is positive, the wave induced drift is larger when the particle is close to the interface as compared to the drift induced when the particle is close to the bottom
surface. For negative (adverse) vorticity in the bottom layer, the largest drift tends to occur at the bottom surface (i.e. the bottom flat bed) when the flow is close to having a stagnation point at the bottom, as also observed in \cite{ribeiro2017flow} for linear shear flow.
\item For flows close to having a stagnation point at the crest, as shown in figure \ref{fig19:paths}, the negative vorticity in the bottom layer clearly increases the drift of the particles in the top irrotational layer relative to the background current, particularly for particles at the surface. 
\item For flows close to having a stagnation point at the bottom, as in figure \ref{fig20:paths}, a stronger negative vorticity in the bottom layer tends to decrease the relative drifts of the particles at different vertical positions in the top layer, as observed by comparing figure \ref{fig19:paths}c and figure \ref{fig20:paths}c.
\end{enumerate}

With a shear layer on an irrotational layer, figure \ref{fig21:paths} shows the particle trajectories for flows close to having a stagnation point at the crest and figure \ref{fig22:paths} demonstrates the particle trajectories for flows close to having a stagnation point at the location when the vorticity jump occurs. Again, we can see from both figures that 
the wave-induced drifts are positive in the direction of the propagating waves. Some conspicuous features are observed as described below:
\begin{enumerate}
\item In the bottom irrotational layer, the drift is always larger when the particle is close to the interface, consistent with particle paths in irrotional flow \citep{paprota2018particle}. 
\item In the top rotational layer, when the vorticity is positive, the wave-induced drift is larger when the particle is near the surface as comapred to the drift when the particle
is close to the interface of the two layers; while with negative vorticity, for flows close to having a stagnation point at the interface, the particle with the largest drift is at the interface.
\item A negative vorticity in the top layer tends to increase the wave-induced drifts of particles in both layers.
\end{enumerate}


\section{Conclusions}\label{sec:con}
In this paper, we investigate the behavior of flow along a new bifurcation curve, and the pressure and particle trajectories under 
waves traveling on a rotational flow with discontinuous vorticity (close to having a stagnation point). For this purpose, we have developed a method for computing the particle paths using the velocity field obtained from a numerical continuation method based on DJ transform of the Euler equations for periodic waves on rotational flow. The continuation method generates a group of waves of fixed relative mass flux until a point where a flow with a stagnation point is reached.

By using the previously mentioned methods, we have recovered new branches of the bifurcation curves not connected to the trivial (laminar) solution, for waves on rotational flow with discontinuous vorticity. Particularly, we show that along the bifurcation curves, the stagnation point can first occur at the position where the vorticity jumps and then again occur at the surface. On the pressure distribution, we show that for the case of shear layer on an irrotational layer, the pressure in the top layer under the crest could be less than the atmosphere pressure and the pressure in the top layer near the interface displays a \lq\lq{plateau}\rq\rq{} behavior when the vorticity is large with a negative sign (i.e., adverse). This \lq\lq{plateau}\rq\rq{} behavior of the pressure has only been observed along the bottom for linear shear flow. We further show the particle trajectories in flows close to having a stagnation point for various cases and qualitatively discuss the influence of vorticity of the rotational layer on the drifts of the particles in both layers.

\section*{Declaration of Interests}
The authors report no conflict of interest.

\section*{Acknowledgments}
The authors are grateful to the anonymous reviewers for their comments and suggestions which helped us improve the paper. L. Chen acknowledges the partial support of the National Natural Science Foundation of China (grant number: 51978506). C. I. Martin acknowledges the support of the Austrian Science Fund (FWF) through research grant P 30878-N32.

\bibliographystyle{jfm}
\bibliography{water-wave-refs}

\end{document}